\documentclass[ALICE,manyauthors]{cernphprep}
\usepackage[comma,square,numbers,sort&compress]{natbib}
\usepackage{hyperref}
\usepackage{lineno}
\usepackage{xspace}
\usepackage{xcolor}
\usepackage{orcidlink}

\begin{document}
%

\newcommand{\pp}           {pp\xspace}
\newcommand{\ppbar}        {\mbox{$\mathrm {p\overline{p}}$}\xspace}
\newcommand{\XeXe}         {\mbox{Xe--Xe}\xspace}
\newcommand{\PbPb}         {\mbox{Pb--Pb}\xspace}
\newcommand{\pA}           {\mbox{pA}\xspace}
\newcommand{\pPb}          {\mbox{p--Pb}\xspace}
\newcommand{\AuAu}         {\mbox{Au--Au}\xspace}
\newcommand{\dAu}          {\mbox{d--Au}\xspace}

\newcommand{\s}            {\ensuremath{\sqrt{s}}\xspace}
\newcommand{\sNN}          {\ensuremath{\sqrt{s_{\mathrm{NN}}}}\xspace}
\newcommand{\pT}           {\ensuremath{p_{\rm T}}\xspace}
\newcommand{\meanpt}       {$\langle p_{\mathrm{T}}\rangle$\xspace}
\newcommand{\ycms}         {\ensuremath{y_{\rm CMS}}\xspace}
\newcommand{\ylab}         {\ensuremath{y_{\rm lab}}\xspace}
\newcommand{\etarange}[1]  {\mbox{$\left | \eta \right |~<~#1$}}
\newcommand{\yrange}[1]    {\mbox{$\left | y \right |~<~#1$}}
\newcommand{\dndy}         {\ensuremath{\mathrm{d}N_\mathrm{ch}/\mathrm{d}y}\xspace}
\newcommand{\dndeta}       {\ensuremath{\mathrm{d}N_\mathrm{ch}/\mathrm{d}\eta}\xspace}
\newcommand{\avdndeta}     {\ensuremath{\langle\dndeta\rangle}\xspace}
\newcommand{\dNdy}         {\ensuremath{\mathrm{d}N_\mathrm{ch}/\mathrm{d}y}\xspace}
\newcommand{\Npart}        {\ensuremath{N_\mathrm{part}}\xspace}
\newcommand{\Ncoll}        {\ensuremath{N_\mathrm{coll}}\xspace}
\newcommand{\dEdx}         {\ensuremath{\textrm{d}E/\textrm{d}x}\xspace}
\newcommand{\RpPb}         {\ensuremath{R_{\rm pPb}}\xspace}

\newcommand{\nineH}        {$\sqrt{s}~=~0.9$~Te\kern-.1emV\xspace}
\newcommand{\seven}        {$\sqrt{s}~=~7$~Te\kern-.1emV\xspace}
\newcommand{\twoH}         {$\sqrt{s}~=~0.2$~Te\kern-.1emV\xspace}
\newcommand{\twosevensix}  {$\sqrt{s}~=~2.76$~Te\kern-.1emV\xspace}
\newcommand{\five}         {$\sqrt{s}~=~5.02$~Te\kern-.1emV\xspace}
\newcommand{\twosevensixnn}{$\sqrt{s_{\mathrm{NN}}}~=~2.76$~Te\kern-.1emV\xspace}
\newcommand{\fivenn}       {$\sqrt{s_{\mathrm{NN}}}~=~5.02$~Te\kern-.1emV\xspace}
\newcommand{\LT}           {L{\'e}vy-Tsallis\xspace}
\newcommand{\GeVc}         {Ge\kern-.1emV/$c$\xspace}
\newcommand{\MeVc}         {Me\kern-.1emV/$c$\xspace}
\newcommand{\TeV}          {Te\kern-.1emV\xspace}
\newcommand{\GeV}          {Ge\kern-.1emV\xspace}
\newcommand{\MeV}          {Me\kern-.1emV\xspace}
\newcommand{\GeVmass}      {Ge\kern-.2emV/$c^2$\xspace}
\newcommand{\MeVmass}      {Me\kern-.2emV/$c^2$\xspace}
\newcommand{\lumi}         {\ensuremath{\mathcal{L}}\xspace}

\newcommand{\ITS}          {\rm{ITS}\xspace}
\newcommand{\TOF}          {\rm{TOF}\xspace}
\newcommand{\ZDC}          {\rm{ZDC}\xspace}
\newcommand{\ZDCs}         {\rm{ZDCs}\xspace}
\newcommand{\ZNA}          {\rm{ZNA}\xspace}
\newcommand{\ZNC}          {\rm{ZNC}\xspace}
\newcommand{\SPD}          {\rm{SPD}\xspace}
\newcommand{\SDD}          {\rm{SDD}\xspace}
\newcommand{\SSD}          {\rm{SSD}\xspace}
\newcommand{\TPC}          {\rm{TPC}\xspace}
\newcommand{\TRD}          {\rm{TRD}\xspace}
\newcommand{\VZERO}        {\rm{V0}\xspace}
\newcommand{\VZEROA}       {\rm{V0A}\xspace}
\newcommand{\VZEROC}       {\rm{V0C}\xspace}
\newcommand{\Vdecay} 	   {\ensuremath{V^{0}}\xspace}

\newcommand{\ee}           {\ensuremath{e^{+}e^{-}}} 
\newcommand{\pip}          {\ensuremath{\pi^{+}}\xspace}
\newcommand{\pim}          {\ensuremath{\pi^{-}}\xspace}
\newcommand{\kap}          {\ensuremath{\rm{K}^{+}}\xspace}
\newcommand{\kam}          {\ensuremath{\rm{K}^{-}}\xspace}
\newcommand{\pbar}         {\ensuremath{\rm\overline{p}}\xspace}
\newcommand{\kzero}        {\ensuremath{{\rm K}^{0}_{\rm{S}}}\xspace}
\newcommand{\lmb}          {\ensuremath{\Lambda}\xspace}
\newcommand{\almb}         {\ensuremath{\overline{\Lambda}}\xspace}
\newcommand{\Om}           {\ensuremath{\Omega^-}\xspace}
\newcommand{\Mo}           {\ensuremath{\overline{\Omega}^+}\xspace}
\newcommand{\X}            {\ensuremath{\Xi^-}\xspace}
\newcommand{\Ix}           {\ensuremath{\overline{\Xi}^+}\xspace}
\newcommand{\Xis}          {\ensuremath{\Xi^{\pm}}\xspace}
\newcommand{\Oms}          {\ensuremath{\Omega^{\pm}}\xspace}
\newcommand{\degree}       {\ensuremath{^{\rm o}}\xspace}

\newcommand{\sF}{\s~=~5.02~TeV\xspace}
\newcommand{\sT}{\s~=~13~TeV\xspace}
\newcommand{\piz}{\ensuremath{\pi^{0}}\xspace}
\newcommand{\e}{\ensuremath{\eta}\xspace}
\newcommand{\Vo}{V\ensuremath{^{0}}\xspace}

\newcommand{\Ngchrec}{\ensuremath{N_{\mathrm{\gamma}}^{\mathrm{\tiny rec}}/ N_{\mathrm{\tiny ch}}^{\mathrm{\tiny rec}}}\xspace}

\newcommand{\Psipairm}{\ensuremath{\Psi_\text{\tiny Pair}}}
\newcommand{\Ng}{\ensuremath{N_{\rm{\gamma}}}\xspace}
\newcommand{\Ngrec}{\ensuremath{N_{\rm{\gamma}}^{\rm{\tiny rec}}}\xspace}
\newcommand{\Ngreci}{\ensuremath{N_{{\gamma,i}}^{\rm{\tiny rec}}}\xspace}
\newcommand{\NgreciX}{\ensuremath{N_{{\gamma,i}}^{\rm{\tiny rec,X}}}\xspace}
\newcommand{\NgreciMC}{\ensuremath{N_{{\gamma,i}}^{\rm{\tiny rec,MC}}}\xspace}

\newcommand{\Ngprod}{\ensuremath{N_{\rm{\gamma}}^{\rm{\tiny prod}}}\xspace}
\newcommand{\Nch}{\ensuremath{N_{\rm{\tiny {ch}}}}\xspace}
\newcommand{\Nchprod}{\ensuremath{N_{\rm{\tiny {ch}}}^{\rm{\tiny prod}}}\xspace}

\newcommand{\Nchrec}{\ensuremath{N_{\rm{\tiny {ch}}}^{\rm{\tiny rec}}}\xspace}

\newcommand{\NchrecX}{\ensuremath{N_{\rm{\tiny {ch}}}^{\rm{\tiny rec,X}}}\xspace}

\newcommand{\Ngch}{\ensuremath{N_{\rm{\gamma}}/N_{\rm{\tiny ch}}}\xspace}

\newcommand{\RConv}{\ensuremath{R_{\rm{\tiny conv}}}}
\newcommand{\ZConv}{\ensuremath{Z_{\rm{\tiny conv}}}}

\newcommand{\qT}           {\ensuremath{q_{\rm T}}\xspace}
\newcommand{\pTmin}           {\ensuremath{p_{\rm T,min}}\xspace}

\begin{titlepage}
\PHyear{2023}       
\PHnumber{043}      
\PHdate{16 March}  


\title{Data-driven precision determination of the material budget in ALICE}

\ShortTitle{Material budget weights}   

\Collaboration{ALICE Collaboration\thanks{See Appendix~\ref{app:collab} for the list of collaboration members}}
\ShortAuthor{ALICE Collaboration} 

\begin{abstract}

The knowledge of the material budget with a high precision is fundamental for measurements of direct photon
production using the photon conversion method due to its direct impact on the total systematic uncertainty. Moreover, it influences many aspects  of the charged-particle reconstruction performance. In this article, two procedures to determine data-driven corrections to the material-budget description in ALICE simulation software are developed.
One is based on the precise knowledge of the gas composition in the Time Projection Chamber. The other is based on the robustness of the ratio between the produced number of photons and charged particles, to a large extent due to the approximate isospin symmetry in the number of produced neutral and charged pions. Both methods are applied to ALICE data allowing for a reduction of the
overall material budget systematic uncertainty from 4.5\% down to
2.5\%. Using these methods, a locally correct material budget is also
achieved. The two proposed methods are generic and can be applied to
any experiment in a similar fashion.
\end{abstract}
\end{titlepage}

\setcounter{page}{2} 


\section{Introduction}
The ALICE experiment is a dedicated heavy-ion experiment at the CERN
LHC~\cite{Aamodt:2008zz,Abelev:2014ffa,ALICE:2022wpn}. In ALICE, photons are
measured using either the calorimeters (PHOS~\cite{ALICE:1999ozr,ALICE:2019cox}, EMCal~\cite{Cortese:1121574,Allen:1272952,ALICE:2022qhn}), or the
Photon Conversion Method (PCM)~\cite{ALICE:2012wos,Abelev:2014ffa},
i.e. via the reconstruction of e$^+$e$^-$ pairs from photon conversions in the detector material. 
The material budget is, expressed in \% of radiation lengths, $X/X_{\mathrm{0}} =$~(11.4 $\pm$ 0.5)\%~\cite{ALICE:2012wos,Abelev:2014ffa}\footnote{Obtained from the geometrical model of the ALICE experiment implemented in the simulation software.}. This value is an average over the pseudorapidity range $|\eta| < 0.9$ and it is integrated in the radial direction ($R$) up to 180~cm, where $R$ is calculated in the transverse $xy$-plane to the beam axis ($z$).
 This uncertainty in the material budget translates into a systematic uncertainty of photon spectrum measurements. For example, direct-photon production\footnote{The terminology used for the different photon sources is as follows. Direct photons: all photons except from neutral meson decays. Thermal photons: photons from the QGP and hadron gas that are dominant at low transverse momentum (\pT$ < 3$\,\GeVc). Prompt photons: produced in hard scatterings (calculable with pQCD) and pre-equlibrium photons that are dominant at high \pT (\pT $> 5$\,\GeVc). } was measured in \PbPb collisions at a center of mass energy per nucleon pair of \sNN~$=$~2.76 TeV in three centrality classes by
ALICE~\cite{Adam:2015lda}. The measurement was done by a combination of
the independent PCM and PHOS measurements.  A low transverse momentum (\pT) excess with
respect to perturbative Quantum Chromodynamics (pQCD) prompt-photon predictions is observed, which can be
attributed to thermal photon emission from 
the quark--gluon plasma (QGP). The current
uncertainties do not allow for discrimination among the proposed theoretical models~\cite{vanHees:2014ida,Paquet:2015lta,Chatterjee:2012dn,Chatterjee:2009rs,Linnyk:2015tha,He:2011zx,Holt:2015cda,Heffernan:2014mla,Turbide:2003si}. For the events in the 0--20\% centrality interval, the low \pT excess is of the order of 10--15\%, and the total uncertainty is approximately 6\% at low \pT with the largest contribution being the 4.5\% of the material budget. Therefore, reducing the uncertainty of the material budget is essential for improving the significance of the direct-photon measurements and, thereby, allowing for a larger discrimination power among the different theoretical models.

The estimated systematic uncertainty related to the material budget of $\pm 4.5$\%~\cite{ALICE:2012wos,Abelev:2014ffa} is an average over
the $R$ range given above. Initially, the systematic uncertainty was estimated by
comparing the reconstructed number of $\gamma$ conversions (\Ng) normalized
to the number of charged particles (\Ngch) between real data (RD) and  Monte Carlo (MC) simulations. Two event generators (PYTHIA~6~\cite{Sjostrand:2006za}
and PHOJET~1.12~\cite{Engel:1995sb}), two secondary vertex
finder algorithms with different optimization criteria, and two momentum ranges
were considered. Local differences of up to 20\% were observed in some
parts of the detector. The development of new procedures to reduce the
systematic uncertainty and to achieve a more accurate local
description of the material budget in the simulation is therefore
mandatory.

This article establishes two data-driven correction methods for a precise determination of the material budget of a given detector using reconstructed photon conversions.
The methods are based either on the existence of a well-known piece of material (e.g. the TPC gas in the ALICE case) or on the robustness of the ratio \Ngch above a low \pT threshold which is largely due to the approximate isospin symmetry in pion production. The two methods are developed for the
ALICE experiment but can, in principle, be employed in any other experiment.

This article is organized as follows. The ALICE experimental setup and
event sample used in this article and the photon reconstruction are
described in \hyperref[sec:det]{Sec.~\ref*{sec:det}}, and
\hyperref[sec:pho]{Sec.~\ref*{sec:pho}}, respectively. The proposed
correction procedures are introduced in
\hyperref[sec:meth]{Sec.~\ref*{sec:meth}}. The results are presented
in \hyperref[sec:res]{Sec.~\ref*{sec:res}}  followed by the conclusions in \hyperref[sec:con]{Sec.~\ref*{sec:con}}.

\section{Detector description and data sample}
\label{sec:det}

A comprehensive description of the ALICE experiment during the LHC Runs 1 and 2 and its performance can be found in Refs.~\cite{Aamodt:2008zz,Abelev:2014ffa}.  The relevant detectors for this analysis are the Inner Tracking System
(ITS)~\cite{Aamodt:2010aa}, the Time Projection Chamber (TPC)~\cite{Alme:2010ke} and the V0 detectors\cite{Abbas:2013taa} which are operated inside a magnetic field up to 
0.5~T directed parallel to the beam axis.  The ITS consists 
of six cylindrical layers of high resolution silicon tracking detectors. 
The two innermost layers located at a radial distance of 3.9\,cm and 7.6\,cm are silicon pixel
detectors (SPD); the two intermediate layers are silicon drift
detectors (SDD) positioned at 15.0 cm and at 23.9\,cm; and the two
outermost layers are silicon strip detectors (SSD) at 38.0\,cm and 43.0\,cm. 
It measures the position of the primary collision vertex, the impact parameter of the tracks, and improves considerably the track \pT resolution at high \pT. The SDD and SSD layers provide the amplitude of the charged signal that is used for particle identification through the measurement of the specific energy loss  (\dEdx).
The TPC is a large ($\sim$ 85 m$^3$) cylindrical drift detector filled
during 2017 with a Ne/CO$_2$/N$_2$ (90/10/5) gas mixture. It covers a
pseudorapidity acceptance of $|\eta|< 0.9$ over the full azimuthal
range, with a maximum of 159 reconstructed space points along the path
of the track. In addition to the space points for the track
reconstruction, the TPC provides particle identification via the
measurement of the \dEdx.  The V0 detector, which is made of
two arrays of 32 plastic scintillators located at 
$2.8 < \eta < 5.1$ (V0A) and $-3.7 <\eta < -1.7$ (V0C), is used for triggering on the collisions~\cite{Abbas:2013taa}.

The analyses presented here use the low intensity part (up to few hundred Hz interaction rate) of the data recorded in 2017 during the LHC pp run at \sF. A total of about 4 $\times$ 10$^7$ pp collisions recorded with a minimum-bias (MB)
trigger are used. The MB trigger was defined by signals in both V0 detectors in coincidence with a
bunch crossing  to minimize the contribution from diffractive interactions.
Contamination from beam-induced background events, produced outside the interaction region, 
is removed using the timing information of the V0 detectors and
taking into account the correlation between tracklets and clusters in
the SPD detector~\cite{Abelev:2014ffa}. The events used for the
analysis are required to have a primary vertex in the fiducial region
$|z| < 10$\,cm along the beam-line direction. The primary vertex is reconstructed 
either using global tracks (with ITS and TPC information) or using SPD tracklets. The contamination from in-bunch pile-up
events is removed offline by excluding events with multiple vertices reconstructed in the SPD~\cite{ALICE:2015olq}.

In general, MC simulations use a geometrical model of the ALICE detectors, an event generator as input, and a particle transport software, GEANT3\cite{Brun:1987ma} in the ALICE case.

\section{Photon reconstruction}
\label{sec:pho}
\begin{figure}[hbt]
\centering
\includegraphics[width=0.49\textwidth]{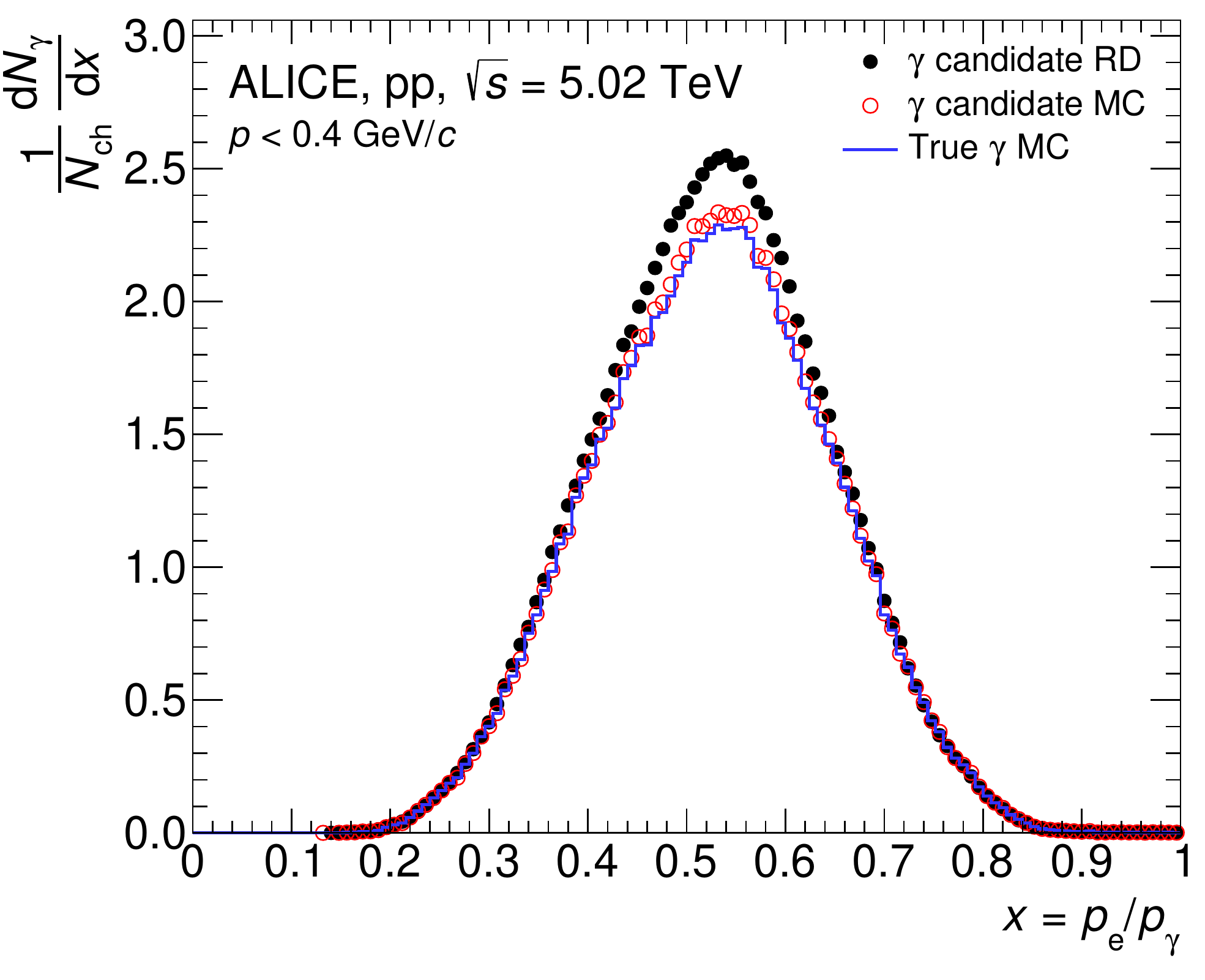}
\includegraphics[width=0.49\textwidth]{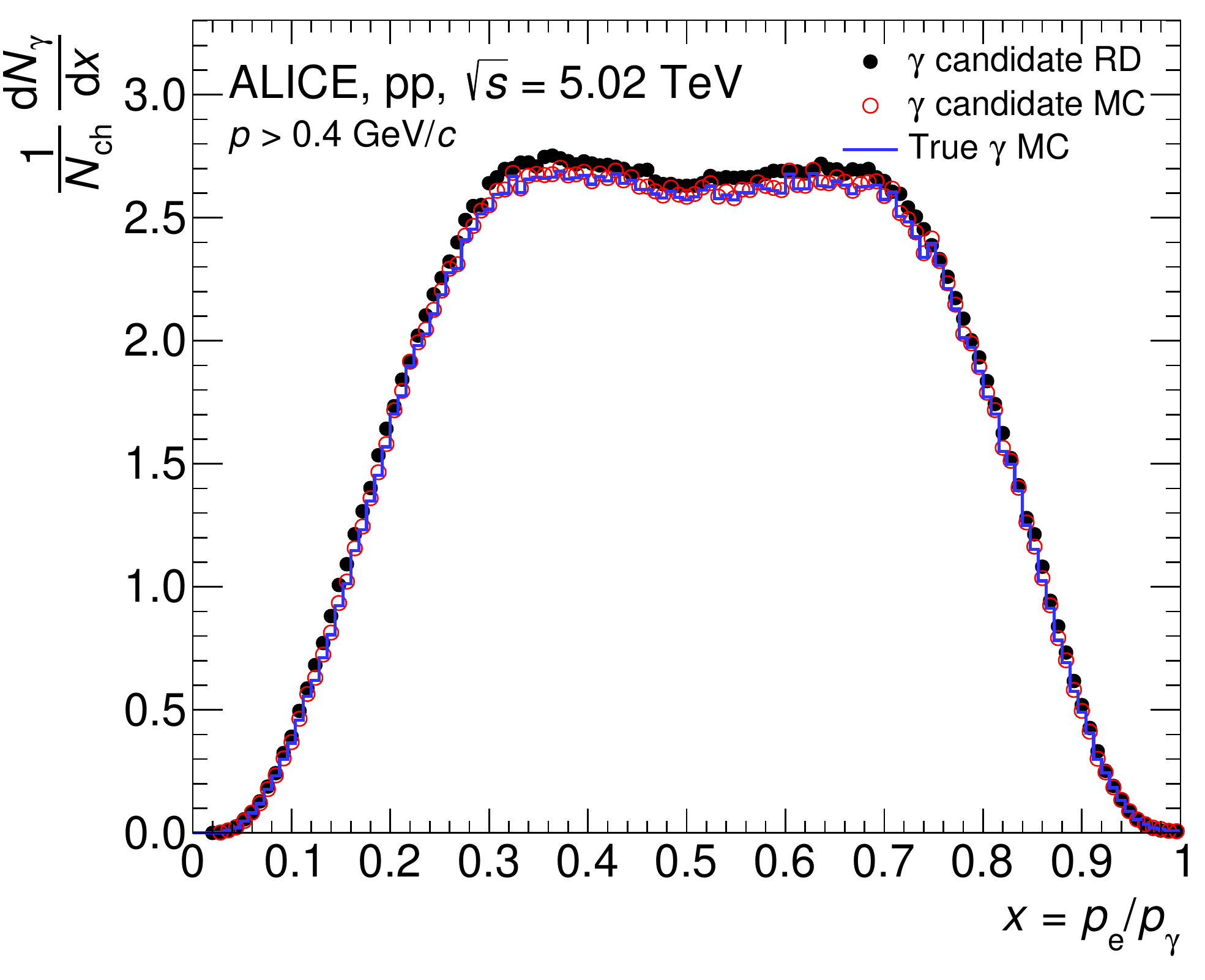}
\caption{Distribution of the electron fractional energy ($x$) for  photon candidates with $p<0.4$\,\GeVc (left) and with $p>0.4$\,\GeVc
  (right) reconstructed in real data (black points), in Monte Carlo simulations (red points), and  for verified photons in MC simulations (blue lines) where PYTHIA8 is used as input event generator.}
\label{fig:photonAsy}
\end{figure}

Photons are reconstructed by measuring the e$^+$e$^-$ pairs produced in
photon conversions in the detector material. Charged tracks are
reconstructed in the ALICE central barrel with the
ITS~\cite{Aamodt:2010aa} and the TPC~\cite{Alme:2010ke}, working together or independently. Two secondary-vertex algorithms with different optimization
criteria are used in this analysis to search for oppositely-charged
track pairs originating from a common (secondary) vertex, referred to
as \Vo ~\cite{Abelev:2014ffa}. The \Vo sample consists mainly of
\kzero, \lmb, \almb decays and $\gamma$ conversions.
Selection criteria based on track quality, particle identification and the topology of a photon conversion are applied. The complete list
of selection criteria is summarized in
\hyperref[tab:cri]{Table~\ref*{tab:cri}}. 
Electrons, positrons, and photons are
required to be within $|\eta| <$ 0.8. 
In order to ensure good track quality, a minimum track transverse momentum of 50 \MeVc and a fraction of TPC clusters
over findable clusters (the number of geometrically possible
clusters which can be assigned to a track) above 0.6 were 
required. The conversion point of the photon candidates
should be inside the \e acceptance and the conversion radius should 
be inside 0~$< R<$ 180 cm and within the limits given by the so called `line cut' (see \hyperref[tab:cri]{Table~\ref*{tab:cri}}) to ensure that the photons come from the primary vertex.
Electron identification and pion rejection are performed by using the specific energy loss \dEdx in the TPC. 
The selection and rejection criteria are based on the 
number of standard deviations ($n\sigma_{e}$ and $n\sigma_{\pi}$ ) around the electron and pion hypothesis, where $\sigma$ is the standard deviation of the energy loss measurement. The remaining contamination from \lmb, \almb, and \kzero is further reduced using a two-dimensional selection in the ($\alpha$, \qT) distribution, known as the Armenteros-Podolanski plot~\cite{podolanski1954iii}; $\alpha$ is the longitudinal
momentum asymmetry of positive and negative daughter tracks, defined as 
$\alpha = (p_\mathrm{L}^+ - p_\mathrm{L}^-)/(p_\mathrm{L}^+ + p_\mathrm{L}^-)$, and \qT is  the transverse momentum of the decay particles with respect to the \Vo momentum.
\begin{table}[htb!]
    \centering
      \caption{Selection criteria applied to the \Vo sample to select photons among the different particles.}
\begin{tabular}{|l |c|}\hline
{\textbf{Track reconstruction }}&  \\
  \cline{1-1}
 e$^\pm$ track \pT       & {$\pT > 0.05$~\GeVc}  \\ 
 e$^\pm$ track $\eta$    & {$|\eta|<0.8$} \\ 
  $N_{\mbox{\tiny clusters}}/N_{\mbox{\tiny findable clusters}}$ & {$>60\%$} \\
   conversion radius      & {$0\,<\, R\,<\,180$\,cm} \\
   line cut                & {$R > |Z| \times ZR_{\mbox{\tiny Slope}} - Z_{0}$}    \\
                        &  {with $ZR_{\mbox{\tiny Slope}} = \tan{(2 \times \arctan(\exp(-\eta_{\mbox{\tiny cut}})))}$}   \\
                        &  {where $Z_{0}~=~$7\,cm, $\eta_{\mbox{\tiny cut}}~=~$0.8}  \\ \hline
 {\textbf{Track identification }} & \\
   \cline{1-1}
  $n \sigma_{\mathrm{e}}$ TPC  & {$-3  < n \sigma_{\mathrm{e}} <5$}                             \\ 
  {$n \sigma_{\pi}$ TPC}      & {$n \sigma_{\pi} >2$ for $0.25<p<3.5$~GeV/$c$}\\
       & $n \sigma_{\pi} >0.5$ for $p>3.5$~GeV/$c$  \\ \hline

  {\textbf{Conversion $\gamma$ topology }} & \\
   \cline{1-1}
 $q_{\mbox{\tiny T}}$        &  {\strut $q_{\mbox{\tiny T}} < 0.05\sqrt{1-(\alpha/0.95)^2}$~\GeVc}\\ 
  photon fit quality     & {$\chi^2_{\mbox{\tiny max}} = 20$}   \\ 
  $\psi_{\rm pair}$  &  {$|\psi_{\rm pair}|< 0.1\,(1 - \chi^2 / \chi^2_{\mbox{\tiny max}})$}        \\ \hline
 
 \end{tabular}
     \label{tab:cri}
\end{table}
Conversion electrons have a preferred emission orientation that can be described by the angle $\psi_{\mbox{\tiny pair}}$ between the plane that is perpendicular to the magnetic field (x-y plane) and the plane defined by the opening angle of the pair.
A selection on $\psi_{\mbox{\tiny pair}}$ 
together with a cut on the photon $\chi^2$ of the Kalman filter fit~\cite{KalmanFilter} further suppresses the contamination from non-photonic \Vo candidates.
Monte Carlo simulations show that a photon purity above 99\% is
achieved  at all transverse momenta except in the vicinity of \pT
$\sim$ 0.3 \GeVc where it decreases to
98\%. \hyperref[fig:photonAsy]{Figure~\ref*{fig:photonAsy}} shows the
probability that a reconstructed electron carries a certain fraction
of the photon energy ($x = {{p_{\rm e}}/{p_{\gamma}}}$) for photon candidates below and above a momentum of 0.4 \GeVc. 
The RD as well
as reconstructed distributions from a MC simulation based on PYTHIA~8 with the Monash 2013 tune
\cite{Sjostrand:2014zea,Skands:2014pea} as input event generator are shown. 
Converted photons with $p$\xspace $>$ 0.4 \GeVc can be reconstructed with electron fractional energies 
from 0 to 1, while at lower $p$ only largely symmetric conversions are
detected. Differences between the data and MC distributions are largely due to
different photon momentum distributions that are not yet equalized at this stage (see \hyperref[eq:Theta]{Eq.~(\ref*{eq:Theta})}).
A high purity in the photon sample can be inferred from the similarity of the red points (MC) and blue curves that represent only MC verified photons (\hyperref[fig:photonAsy]{Fig.~\ref*{fig:photonAsy}}).

\begin{figure}[hbt]
\centering 
\includegraphics[width=0.65\textwidth]{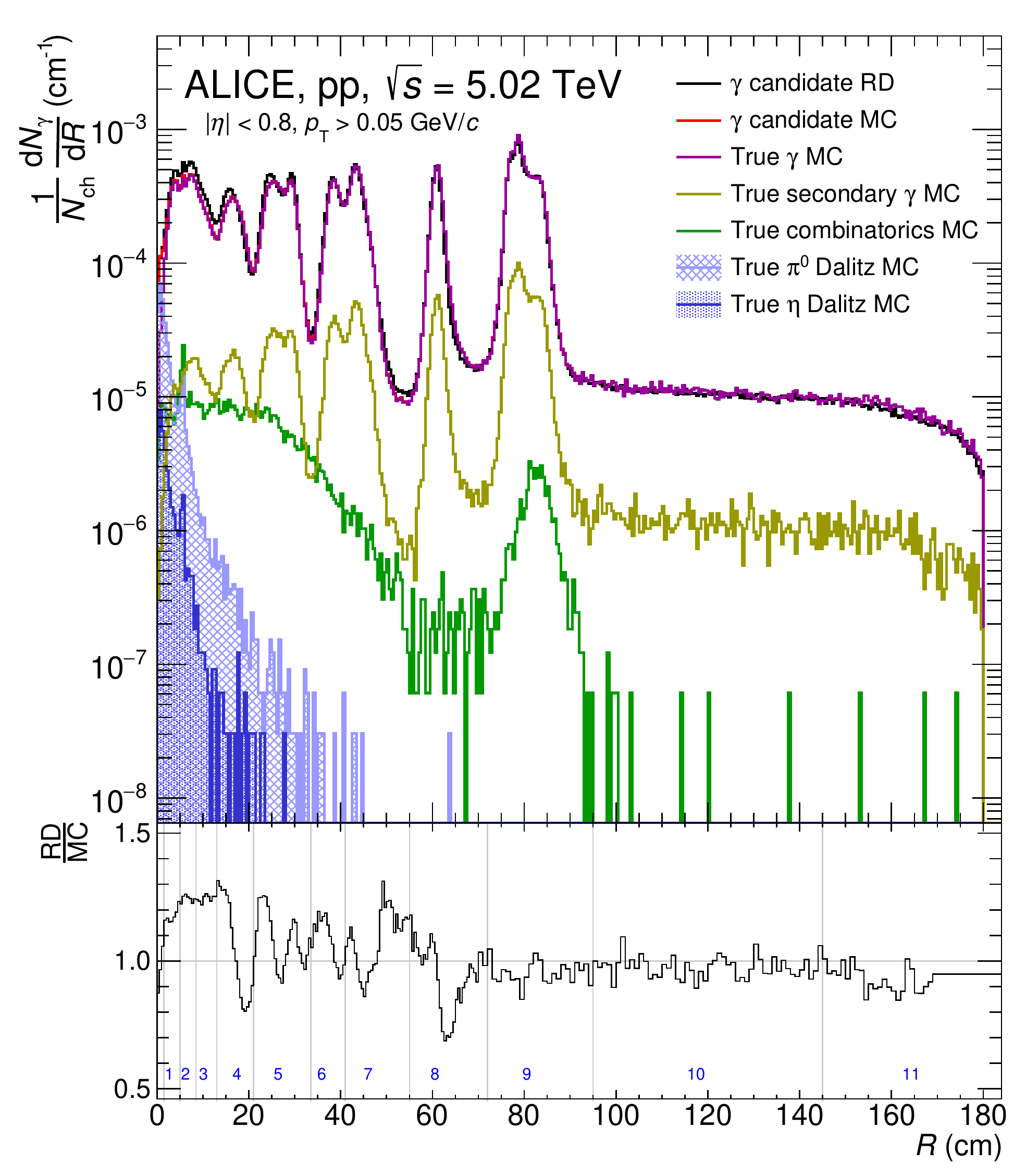}
\caption{(Top) Radial distributions of reconstructed photon vertices in experimental data  and in MC simulations are shown as black and red lines, respectively. Using the full MC information contributions from true photons (purple line), secondary photons (olive line), and combinatorial pairs (green line) are identified. The e$^+$e$^-$ pairs from \piz (or $\eta$) Dalitz decays, wrongly identified as photon  conversions, are depicted as blue shaded areas. 
 (Bottom) Ratio of the radial distribution of reconstructed photon vertices in  RD (black line) and MC (red line).  Vertical lines in grey and blue numbers indicate the twelve radial intervals and their respective indices used in the analysis.}
\label{fig:photonR}
\end{figure}

\hyperref[fig:photonR]{Figure~\ref*{fig:photonR}} displays the radial
distribution of photon conversion vertices. The experimental data as well
as reconstructed distributions from a MC simulation based on PYTHIA~8 with the Monash 2013 tune
\cite{Sjostrand:2014zea,Skands:2014pea} as input event generator are shown. The radial distribution of reconstructed conversion vertices clearly reveals the different detector structures corresponding to the ITS and TPC. The experimental distribution is compared to the one obtained from MC simulations that
accounts for the time-dependent variations of the detector conditions.
The main goal is to select  the primary photon sample, i.e. photons coming from electromagnetic decays of neutral mesons or direct photons. 
Additionally, there are three types of background contributions shown in the figure that need special treatment. i) Primary e$^+$e$^-$ pairs from \piz (or $\eta$) Dalitz decays wrongly detected as conversion photons, mainly localized at radii smaller than 5 cm, can be suppressed by a minimum cut of 5 cm in the analysis. ii) Random combinatorial background, which
is subtracted both in RD and MC based on the MC. 
iii) a 5--10\% contribution of secondary photons~\cite{ALICE-PUBLIC-2017-005} from weak decays of \kzero (e.g. \kzero$\rightarrow$\piz\piz) and \lmb (\almb) (e.g. \lmb$\rightarrow$n\piz) hadrons, and interactions in the detector material. 
The contribution from weak decays is estimated in the data by using a particle decay simulation called “cocktail simulation”~\cite{Acharya:2018dqe} based on parametrizations of measured particle spectra, and in MC using the full MC information (labelled `True' in \hyperref[fig:photonR]{Fig.~\ref*{fig:photonR}} ). The contribution from interactions in the detector material is taken from MC, for both MC and real data (RD). The total secondary contribution is then subtracted from the photon sample in the data and in the MC simulated sample for each radial interval.

\section{Calibration methods}
\label{sec:meth}
The  RD to MC comparison of the number of reconstructed photons as a function of the conversion point radius shown
in \hyperref[fig:photonR]{Fig.~\ref*{fig:photonR}} reveals local differences of up to 20\% in the number of reconstructed photons\footnote{This differences could not be reduced further by checks of the ALICE geometrical model}. This provides evidence that the implementation of the material budget in the geometrical model of ALICE is not accurate enough in some parts of the detector, which may impact the precision of various analyses in ALICE.

Two data-driven calibration methods (the $\omega_i$ and ${\Omega}_i$ calibration weights, \hyperref[sec:Methodomega]{Sec.~\ref*{sec:Methodomega}}, \hyperref[sec:MethodOmegaTilde]{Sec.~\ref*{sec:MethodOmegaTilde}} ) were developed in order to mitigate the
differences by correcting the detector material description in MC simulations, and thus,
reducing the systematic uncertainty. The complete radial range is subdivided in twelve intervals
as shown in \hyperref[fig:photonR]{Fig.~\ref*{fig:photonR}}.
The resulting calibration weights
are then used to scale the photon reconstruction efficiency as follows:

\begin{equation}
\epsilon_\gamma^\mathrm{MC,corr}(p_\mathrm{T}) =  \frac{ \sum\nolimits_{i} 
W_i \times \mathrm{d}N_{\gamma,i}^\mathrm{rec,MC}/\mathrm{d}p_\mathrm{T}  } 
{\mathrm{d}N_\gamma^\mathrm{prod}/\mathrm{d}p_\mathrm{T}},
\label{eq:efi}
\end{equation}
where $W_\mathrm{i}$ are the correction factors in each radial interval $i$ ($W_i\equiv\omega_i$ or 
$W_i\equiv{\Omega_i}$),
d\NgreciMC/d\pT  is the number of primary photons reconstructed in the Monte
Carlo simulated data at a given \pT and radial interval~$i$ (see \hyperref[fig:photonR]{Fig.~\ref*{fig:photonR}}) in the pseudorapidity range $|\eta|<0.8$, and d\Ngprod/d\pT is the total number of photons produced at a given \pT as given by the  input event generator used in the MC within the same pseudorapidity interval.

Another approach for applying the calibration weights is to scale the density of the detector materials used in the geometrical model of the detector by the correction factors  $W_\mathrm{i}$, and produce new MC simulations. While the method given by \hyperref[eq:efi]{Eq.~\ref*{eq:efi}} is only valid for analyses involving photons, the scaling of the density is valid for all analyses. The only disadvantage is the need of creating new simulation samples, with the corresponding CPU needs. 

\subsection{TPC-gas based calibration weights: \texorpdfstring{$\omega_i$}{wi}\xspace}
\label{sec:Methodomega}

The material-budget correction via $\omega_i$ calibration weights
exploits the fact that a well-known and ho\-mo\-ge\-neous part of the detector
material can be used as a reference to calibrate the material budget of the rest of the detector known with less precision.
The TPC gas volume~\cite{Alme:2010ke} in the fiducial range used for
photon reconstruction, $95 <  R < 145$\,cm, is perfectly suited for this purpose. 
The TPC gas material budget 
depends on the exact chemical composition, temperature 
and pressure of the gas. During data taking, variations in the TPC gas composition and pressure are monitored with
a gas chromatograph and a pressure gauge, respectively, and
applied accordingly in the MC
simulations via access to the conditions and calibrations database with a granularity of few hours. The temperature gradients inside the TPC are controlled
to a root-mean-square (rms) deviation of less than $0.05\,\degree$C~\cite{Alme:2010ke} in order to control the drift properties. With the TPC gas monitoring system~\cite{Alme:2010ke} the TPC gas density is known to the per mil level.
The TPC-gas based calibration weights ($\omega_i$ ) are then given by
\begin{equation}
\omega_i  = \frac{{N_{\mathrm{\gamma},i}^{\mbox{\tiny rec,RD}}}/  {N_{\gamma,\mathrm{gas}}^{\mbox{\tiny rec,RD}}}} {{N_{\mathrm{\gamma},i}^{\prime\,\mbox{\tiny rec,MC}}}/{N_{\gamma,\mathrm{gas}}^{\prime\,\mbox{\tiny rec,MC}}}}, 
\label{eq:wi}
\end{equation}

\noindent
where \NgreciX is the number of reconstructed primary  photons in a given radial interval $i$ (denoted by `gas' for the reference one) expressed as

\begin{equation}
N_{\mathrm{\gamma},i}^{\mbox{\tiny rec,X}} =  \int^\infty_{p_\mathrm{{T,\mathrm{min}}}}
\Tilde{P}_i^{\mbox{\tiny X}} (p_\mathrm{T}) \times \Tilde{\epsilon}_i^{\mbox{\tiny X}}(p_\mathrm{T}) \frac{dN_{\mathrm{\gamma}}^\mathrm{\mbox{\tiny prod,X}}}{dp_\mathrm{T}} dp_\mathrm{T}
\equiv
P_i^{\mbox{\tiny X}} \times \epsilon_{\mathrm{\gamma},i}^{\mbox{\tiny X}} \times 
N_\gamma^{\mbox{\tiny prod,X}},
\label{eq:ngrecprodprobeff}
\end{equation}

where X refers both to RD and to MC,
$p_{\mbox{\tiny T,min}} = 0.05$ \GeVc, $N_{\gamma}^{\mbox{\tiny prod}}$ is the number of produced primary photons either in RD or in MC. The  photon conversion probability, and
the photon reconstruction efficiency in a given radial interval $i$ are denoted by $P_{i}$ and
$\epsilon_{\gamma,i}$, respectively. 
The reweighted $N_{\gamma,i}^{\prime\,\mbox{\tiny rec}} $ is defined later in \hyperref[eq:Ngprime]{Eq.~(\ref*{eq:Ngprime})}.
The photon reconstruction efficiency is calculated using MC simulations. The RD and MC labels are also used in the efficiency to emphasize possible differences of the RD efficiency compared to the values obtained in the MC simulations.
The ratio between two radial intervals suppresses the impact of different numbers of produced
photons, or the overall reconstruction efficiency, between data and Monte
Carlo simulations.


\subsection{Pion-isospin-symmetry based calibration weights: \texorpdfstring{${\Omega}_i$}{wi}\xspace } 

\label{sec:MethodOmegaTilde}
The material budget correction via the ${\Omega}_i$ calibration
weights exploits the robustness of the ratio of the
number of reconstructed photons to the number of reconstructed charged
particles
(\Ngchrec) above a certain low \pT threshold. The reason for the robustness is the approximate isospin symmetry~\cite{Fritzsch:2005zza} in the number of produced charged and neutral pions, and that charged pions and photons from \piz decays are the dominant contributions to the number of charged particles (90\% of charged particles are charged pions)~\cite{Acharya:2019yoi,Acharya:2018qsh} and the total number of photons~\cite{Acharya:2018dqe}, respectively. 'Approximate' is used to point out that electromagnetic decays of the \e , $\omega$ and $\eta^\prime$ mesons are  cases that violate the pion isospin symmetry.
By employing PYTHIA~8 and PHOJET~\cite{Engel:1995sb} event generators it was checked that the ratio is constant
at the per mil level even if the charged-particle multiplicity or the
photon multiplicity differ by 10--20\% depending on the collision energy and event generator.
In summary, the ratio of the number of reconstructed primary photons 
in a given radial interval (\Ngreci ) divided by the number of reconstructed primary charged
particles ($N_{\mathrm{ch}}$) in RD over the same quantity in MC (${\Omega}_i$) is sensitive to the correctness of the
detector material implementation; thus, the ${\Omega}_i$
can be used as calibration weights. 

The pion-isospin-symmetry based
calibration weights (${\Omega}_i$) are then defined as
\begin{equation}
{\Omega}_i = \frac{{N_{\gamma,i}^{\mbox{\tiny rec,RD}}}/{N_{\mbox{\tiny ch}}^{\mbox{\tiny rec,RD}}}} 
{{N_{\gamma,i}^{\prime,\mbox{\tiny rec,MC}}}/{N_{\mbox{\tiny ch}}^{\mbox{\tiny rec,MC}}}}, 
\label{eq:OmegaTildei}
\end{equation}

\noindent 
where (\NchrecX) is the number of
reconstructed primary tracks with a transverse momentum \pT$>$
0.15~\GeVc in the pseudorapidity range $|\eta|<0.8$, X refers to RD or to MC, and
$N_{\gamma,i}^{\mbox{\tiny rec}}$ is the number of reconstructed primary photons in the radial interval $i$ with transverse momentum above a minimum value of 0.05\,\GeVc (see \hyperref[eq:ngrecprodprobeff]{Eq.~(\ref*{eq:ngrecprodprobeff})}).
The reweighted $N_{\gamma,i}^{\prime\,\mbox{\tiny rec}} $ is defined later in \hyperref[eq:Ngprime]{Eq.~(\ref*{eq:Ngprime})}. Charged tracks are selected with selections on the number of space points
used for tracking and on the quality of the track fit, as well as on the distance of closest approach to the reconstructed vertex~\cite{Acharya:2018qsh}.
The contribution of secondary charged particles is subtracted
in the case of data using the measurement performed in ALICE in the same data set~\cite{Acharya:2018qsh}, and in MC using the full MC information.
The normalization to
$N_{\mathrm{ch}}$ minimizes the impact of the model dependence of a
given inclusive photon production yield $N^{\mathrm{prod}}_{\gamma}$
in an event generator relative to RD.

\subsection{Comparison of the two calibration methods}
\label{sec:Omega2omega}
The comparison of the two calibration factors described in
\hyperref[sec:Methodomega]{Sec.~\ref*{sec:Methodomega}} and
\hyperref[sec:MethodOmegaTilde]{Sec.~\ref*{sec:MethodOmegaTilde}}
carries very valuable information.  In order to gain insights into potential differences between the two sets of correction factors, it is useful to write them in terms of the conversion probability $P_i$ in a given radial interval.

\noindent
Taking into account
\hyperref[eq:ngrecprodprobeff]{Eq.~(\ref*{eq:ngrecprodprobeff})}, the
$\omega_i$ calculation given by \hyperref[eq:wi]{Eq.~(\ref*{eq:wi})}
transforms into

\begin{equation}
\omega_i=\frac {{P_{i}^{\mbox{\tiny RD}} \times \epsilon_{\gamma,i}^{\mbox{\tiny RD}} }} 
 {{P_{i}^{\mbox{\tiny MC}} \times \epsilon_{\gamma,i}^{\mbox{\tiny MC}}    }}             \times
\frac{{P_{\mbox{\tiny gas}}^{\mbox{\tiny MC}} \times \epsilon_{\gamma,\mbox{\tiny gas}}^{\mbox{\tiny MC}}}}
{{P_{\mbox{\tiny gas}}^{\mbox{\tiny RD}} \times \epsilon_{\gamma,\mbox{\tiny gas}}^{ \mbox{\tiny RD}} }}.
\label{eq:wia}
\end{equation}

\noindent
Under the assumption that the gas is a well-known material, the
conversion probabilities in MC and RD agree for the reference radial interval, i.e.

\begin{equation}
  P_{\mbox{\tiny gas}}^{\mbox{\tiny RD}} = P_{\mbox{\tiny gas}}^{\mbox{\tiny MC}}. 
\label{eq:gasknown}
\end{equation}

\noindent
Then, the $\omega_i$ calculation given by
\hyperref[eq:wia]{Eq.~(\ref*{eq:wia})} reduces to

\begin{equation}
\omega_i = \frac{P_{i}^{\mbox{\tiny RD} } \times \epsilon_{\gamma,i}^{\mbox{\tiny RD}} \times \epsilon_{\gamma,\mbox{\tiny gas}}^{ \mbox{\tiny MC}} } {{P_{i}^{\mbox{\tiny MC}}\times \epsilon_{\gamma,i}^{\mbox{\tiny MC}}} \times \epsilon_{\gamma,\mbox{\tiny gas}}^{\mbox{\tiny RD}}}, 
\label{eq:omegai}
\end{equation}

For the pion-isospin-symmetry based calibration method, 
${\Omega}_i$, the number of
reconstructed primary charged particles (\Nch) in the same $|\eta|$
range is also needed:
\begin{equation}
N_{\mbox{\tiny ch}}^{\mbox{\tiny rec,X}} =
N_{\mbox{\tiny ch}}^{\mbox{\tiny prod,X}} 
\times \epsilon_{\mbox{\tiny track}}^{\mbox{\tiny X}},  
\label{eq:Nchtr}
\end{equation}
where X refers to RD or MC.

Using  \hyperref[eq:OmegaTildei]{Eq.~(\ref*{eq:OmegaTildei})}   and 
\hyperref[eq:Nchtr]{Eq.~(\ref*{eq:Nchtr})} the $\Omega_i$ are given by
\begin{equation}
{\Omega}_i 
= \frac{P_{i}^{\mbox{\tiny RD}} \times \epsilon_{\gamma,i}^{\mbox{\tiny RD}} \times \epsilon_{\mbox{\tiny track}}^{\mbox{\tiny MC}} \times
N_{\gamma}^{\mathrm{prod,RD}}/
N_{\mathrm{ch}}^{\mathrm{prod,RD}} }
{ {P_{i}^{\mbox{\tiny MC}}} \times {\epsilon_{\gamma,i}^{{\mbox{\tiny MC}}}} \times \epsilon_{\mbox{\tiny track}}^{ \mbox{\tiny RD}} \times
N_{\gamma}^{\mathrm{prod,MC}}/N_{\mathrm{ch}}^{\mathrm{prod,MC}}}.
\label{eq:OmegaTildeibNch}
\end{equation}
\noindent
By employing the same MC simulations with PYTHIA~8 and PHOJET as event generators it was verified that 
the quantity \Ngprod/\Nchprod is constant within approximately 1.5\% when varying $p_{\mathrm{T,min}}$ between 0.15 \GeVc and 0.25 \GeVc.  The $p_{\mathrm{T,min}}$ value reduces the diffractive contribution that is different among the two event generators. In this work, the value for the ratio
\Ngprod/\Nchprod obtained from PYTHIA is assumed for RD and differences between PYTHIA and PHOJET are taken as part of the systematic uncertainties.
 With this assumption for the quantity \Ngprod/\Nchprod, the calibration weights ${\Omega}_i$ 
reduce to
\begin{equation}
{\Omega}_i 
= \frac{P_{i}^{\mbox{\tiny RD}} \times \epsilon_{\gamma,i}^{\mbox{\tiny RD}} \times \epsilon_{\mbox{\tiny track}}^{\mbox{\tiny MC}} }{ {P_{i}^{\mbox{\tiny MC}}} \times {\epsilon_{\gamma,i}^{{\mbox{\tiny MC}}}} \times \epsilon_{\mbox{\tiny track}}^{ \mbox{\tiny RD}} }.
\label{eq:OmegaTildeib}
\end{equation}
\noindent
In case a constant (\pT and $R$-independent) factor between the \Vo reconstruction efficiencies in RD and MC would exist 
($\epsilon^{\mbox{\tiny RD}}_{\gamma,i}/\epsilon^{\mbox{\tiny MC}}_{\gamma,i}$) it would drop out for the $\omega_i$ weights (see \hyperref[eq:wia]{Eq.~(\ref*{eq:wia})}). According to 
\hyperref[eq:OmegaTildeib]{Eq.~(\ref*{eq:OmegaTildeib})} this is not the case for ${\Omega}_i$ weights.

Using \hyperref[eq:wia]{Eq.~(\ref*{eq:wia})} or \hyperref[eq:omegai]{Eq.~(\ref*{eq:omegai})} and \hyperref[eq:OmegaTildeib]{Eq.~(\ref*{eq:OmegaTildeib})}, the ratio
${\Omega}_i$/$\omega_i$ is given by

\begin{equation}
\frac{{\Omega}_i}{\omega_i} = \frac{ P_{\mbox{\tiny gas}}^{\mbox{\tiny RD}} \times \epsilon_{\gamma,\mbox{\tiny gas}}^{\mbox{\tiny RD} } \times \epsilon_{\mbox{\tiny track}}^{ \mbox{\tiny MC}} }
{ P_{\mbox{\tiny gas}}^{\mbox{\tiny MC}} \times \epsilon_{\gamma,\mbox{\tiny gas}}^{\mbox{\tiny MC}} \times \epsilon_{\mbox{\tiny track}}^{ \mbox{\tiny RD}} },
\label{eq:OmegaToomegaratioWithP}
\end{equation}

or
\begin{equation}
\frac{{\Omega}_i}{\omega_i} = \frac{ \epsilon_{\gamma,\mbox{\tiny gas}}^{\mbox{\tiny RD} } \times \epsilon_{\mbox{\tiny track}}^{\mbox{\tiny MC}}}{\epsilon_{\gamma,\mbox{\tiny gas}}^{\mbox{\tiny MC}} \times \epsilon_{\mbox{\tiny track}}^{\mbox{\tiny RD}}}.
\label{eq:OmegaToomegaratio}
\end{equation}

\noindent

By calculating the ${\Omega}_i$/$\omega_i$ ratio one can cross-check if the ratio  $\epsilon_{\gamma,\mbox{\tiny gas}}/\epsilon_{\mbox{\tiny track}}$, i.e. the reconstruction efficiency in the
reference radial interval `gas' over the charged
particle reconstruction efficiency, is reproduced in MC.
In case ${\Omega}_i$/$\omega_i$ $\neq$\, 1, the $\omega_i$ calibration weights cannot be used directly as correction factors.

\section{Determination of \texorpdfstring{$\omega_i$}{wi} and \texorpdfstring{${\Omega}_i$}{Wi}}
\label{sec:res}

All quantities that are needed to compute the $\omega_i$
and ${\Omega}_i$ correction factors are introduced in
\hyperref[sec:meth]{Sec.~\ref*{sec:meth}}.
The calculation of the $\omega_i$ and ${\Omega}_i$  weights follows an
iterative procedure because, as the photon reconstruction efficiency is different for the various radial intervals, the differences between the reconstructed number of photons in RD and in MC can also result from a deviation of the shape of the MC photon \pT spectrum from RD. In order to take this effect into account,
the shape of the MC simulated photon transverse
momentum spectrum is adjusted to the measured shape in RD by applying weights ($\Theta(p_{\mbox{\tiny T}})$)
to the MC simulated data as
\begin{equation}
 \mathrm{d}N{^{\prime\,}}_{\gamma}^{\mbox{\tiny rec,MC}}/\mathrm{d}p_{\mbox{\tiny T}}  =
\mathrm{d}N_{\gamma}^{\mbox{\tiny rec,MC}}/ \mathrm{d}p_{\mbox{\tiny T}} \times \Theta(p_{\mbox{\tiny T}}),  
\label{eq:Ngprime}
\end{equation}
where $\Theta(p_{\mbox{\tiny T}})$ is defined as

\begin{equation}
\Theta (p_{\mbox{\tiny T}}) = 
\frac{N_{\gamma}^{\mbox{\tiny rec,MC}}}{N_{\gamma}^{\mbox{\tiny rec,RD}}} \times \frac{\mathrm{d}N_{\gamma}^{\mbox{\tiny rec,RD}}/\mathrm{d}p_{\mbox{\tiny T}}}
{\mathrm{d}N_{\gamma}^{\mbox{\tiny rec,MC}}/\mathrm{d}p_{\mbox{\tiny T}}}.
\label{eq:Theta}
\end{equation}

\noindent
The total number of photons is preserved as only the shape of the spectrum is modified, 
\begin{equation}
N{^{\prime\,}}_{\gamma}^{\mbox{\tiny rec,MC}}=N_{\gamma}^{\mbox{\tiny rec,MC}}.
\end{equation}
The first step is to align the shape of the reconstructed
transverse momentum distributions of the MC to the  data (\hyperref[eq:Ngprime]{Eq.~(\ref*{eq:Ngprime})}) using the $\Theta(p_{\mbox{\tiny T}})$ factors given in \hyperref[eq:Theta]{Eq.~(\ref*{eq:Theta})}.
This step is performed twice to achieve good agreement. A set of $\omega_i$ and
${\Omega}_i$ is then obtained using \hyperref[eq:wi]{Eq.~(\ref*{eq:wi})} and \hyperref[eq:OmegaTildei]{Eq.~(\ref*{eq:OmegaTildei})}, respectively.
Applying the ${\Omega}_i$ calibration weights results in a
modification of the reconstructed transverse momentum
distribution. Therefore, a second iteration of the complete procedure is
performed, i.e. a new set of $\Theta$(\pT) and
${\Omega}_i$ is computed. Applying this new set of ${\Omega}_i$ does not introduce any further change in the transverse momentum distribution, i.e. the procedure of evaluating $\omega_i$ and ${\Omega}_i$ weights converged.

Four combinations of input event generators and \Vo finders are tested and used for the evaluation of the systematic uncertainties. 
PYTHIA 8 with the Monash 2013 
tune~\cite{Sjostrand:2014zea,Skands:2014pea} and PHOJET  1.12~\cite{Engel:1995sb}, available within the DPMJET 3.0~\cite{Roesler:2000he} package, are used.
The default combination is using PYTHIA 8 as event generator, since the Monash tune is the result of an optimization for the LHC data. For PHOJET  
larger differences as compared with PYTHIA 8 are observed in the simulated charged-particle multiplicity distributions compared to experimental data.
Both event generators show differences in the transverse momentum spectrum with respect to experimental data.
These differences are considered as part of the systematic uncertainties of the resulting material budget weights. The two \Vo finders are called  ``on-the-fly'' and ``offline''. The ``on-the-fly'' \Vo finder searches for \Vo candidates 
during the tracking procedure, when the complete detector information, down to reconstructed clusters is available. The ``offline'' \Vo finder searches for \Vo candidates based on reconstructed tracks, which includes their full momentum vector and uncertainty covariance matrix, but no cluster level information.
Each method results in a somewhat different performance in terms of reconstruction efficiency at different radii. 
The ``on-the-fly'' \Vo finder is the default choice for the calculation of the $\Omega_i$ and $\omega_i$ calibration weights, mostly because of its significantly larger efficiency, and because the photon momenta are calculated at the conversion point.
By varying the  \Vo finder, uncertainties on the reconstruction efficiency and its \pT and $R$ dependence, track selection criteria, and the secondary-vertex algorithm in itself are included.
By varying the event generator, uncertainties on the robustness of  \Ngch are included.  
An additional variation  of $p_{\mathrm{T,min}}$ 
from 0.05\,\GeVc up to 0.2 \GeVc does not yield a sizeable difference in either of the methods.

The systematic uncertainties of the weights ${\Omega}_i$ are calculated according to

\begin{equation}
    \sigma_{\Vo\,\mathrm{finder}} = |{W}_i^\text{PYTHIA,on-the-fly}-{W}_i^\text{PYTHIA,offline} |,
\end{equation}

\begin{equation}
\sigma_{\mathrm{generator}}= | {W}_i^\text{PYTHIA,on-the-fly}-{W}_i^\text{PHOJET,on-the-fly}|,
\end{equation}

\begin{equation}
\sigma_{\rm sys}^2({W}_i)=\sigma_{\Vo\,\mathrm{finder}}^2({W}_i)+\sigma_{\mathrm{generator}}^2({W}_i),
\end{equation}

where $W_i\equiv\omega_i $ or $W_i\equiv{\Omega}_i $.
\hyperref[tab:sysSources]{Table~\ref*{tab:sysSources}} shows details for two selected radial intervals.

\begin{table}
\begin{center}
\caption{Sources of the relative systematic uncertainties (\%) of $\Omega_i$ and $\omega_i$ for two radial intervals. The \Vo finder category includes uncertainties on the reconstruction efficiency and its \pT and $R$ dependence, track selection criteria, as well as two \Vo finder methods. The generator category includes uncertainties on the robustness of  \Ngch. The total systematic uncertainty is given in the last row. }
\begin{tabular}{|l|c|c||c|c|}\hline
& \multicolumn{2}{c||}{$\Omega_i$} & \multicolumn{2}{c|}{$\omega_i$}  \\ \cline{2-5}
& 5\,cm $<R<$ 8.5\,cm  &  95\,cm $<R<$ 145\,cm & 8.5\,cm $<R<$ 13\,cm  &  72\,cm $<R<$ 95\,cm  \\ \hline
{\bf \Vo finder} &  2.74 \% &  2.9\%  &  2.2\% &  1.83\%\\ \hline 
{\bf Generator} & 0.16\%& 2.9\%  & 3.2 \%  &   0.62 \%\\   \hline
\pTmin  & Negligible &  Negligible& Negligible & Negligible\\ \hline \hline
$\sigma_{\rm sys}$ & 2.74\%  &  4.1\% & 3.8\% &  1.93\%\\  \hline
\end{tabular}
\end{center}
\label{tab:sysSources}
\end{table}

The radially averaged weight $\langle{\Omega}\rangle$ is given by
\begin{equation}
\langle{\Omega}\rangle =  \sum_{i} \frac{N_\gamma^{\rm i}}{ N_\gamma^{\rm total}} \times {\Omega}_i,
\end{equation}
and, assuming that the statistical uncertainties are negligible and the systematic uncertainties are correlated among the different radial intervals, its uncertainty $\sigma({\Omega})$ is given  by 
\begin{equation}
\sigma^{}({\Omega}) =  \sum_{i} \frac{N_\gamma^{\rm i}}{ N_\gamma^{\rm total}} \times \sigma^{}({\Omega}_i).
\end{equation}

\begin{figure}[hbt]
\centering
\includegraphics[width=0.55\textwidth]{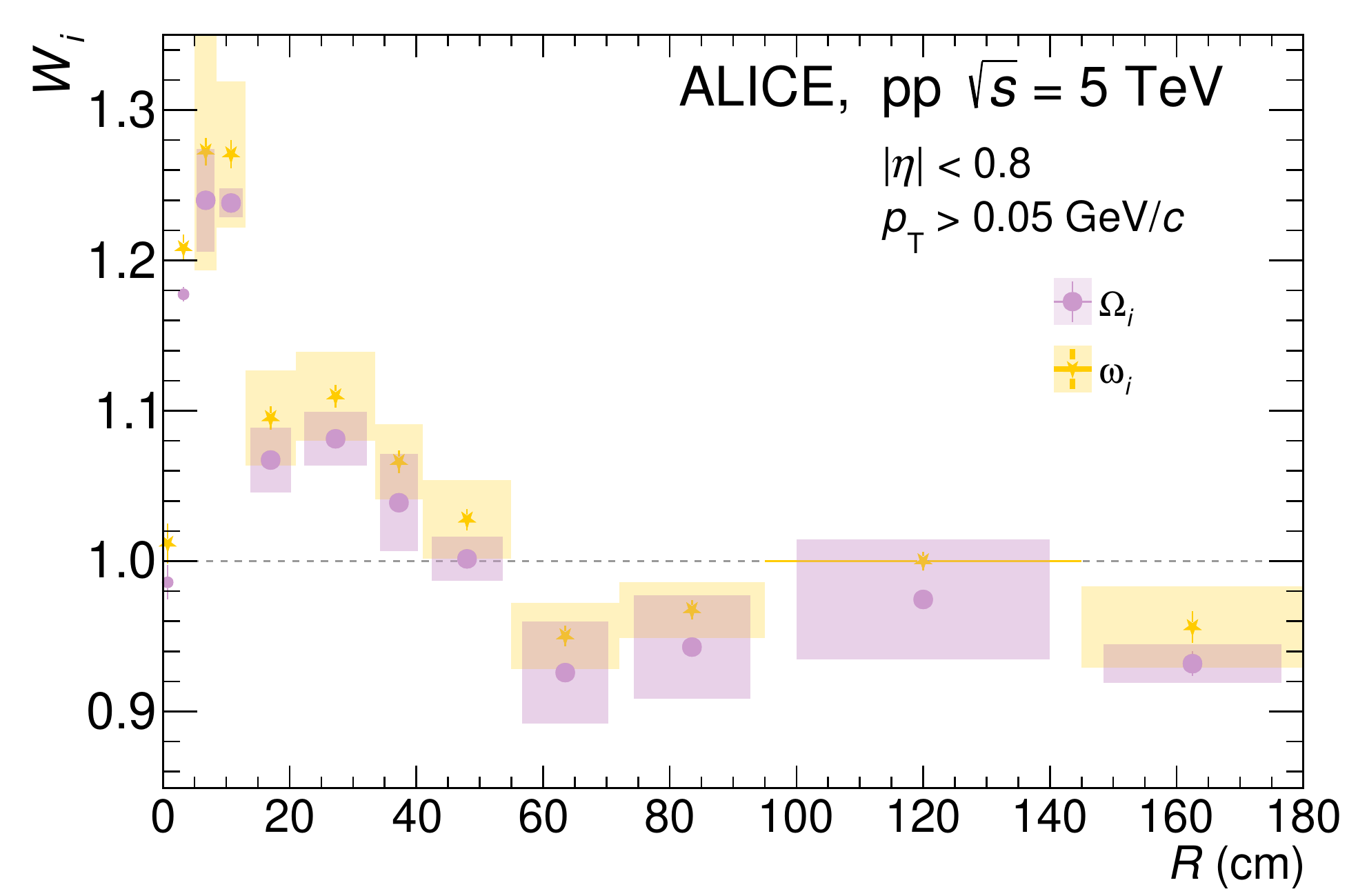}
\caption{Distribution of the weights ${\Omega}_i$  and $\omega_i$ as a function of the radial position of the conversion point. Statistical uncertainties are given by the vertical bars, and systematic uncertainties are depicted by shaded areas, except for the first two intervals where only statistical uncertainties are quoted. Notice the zero $\omega_i$ systematic uncertainty in the calibration intervals (95~cm$< R< $ 145~cm), thus
an horizontal line is drawn displaying the radial interval. The width in spatial direction of the uncertainty box for the $\Omega_i$ was shrunk for better visibility.}
\label{fig:TildeOmega}
\end{figure}

The final set of $\omega_i$ and ${\Omega}_i$ calibration weights is presented in \hyperref[fig:TildeOmega]{ Fig.~\ref*{fig:TildeOmega}}. The calibration weights range from a minimum value of $0.926 \pm 0.034$ to a maximum of $1.240 \pm 0.034$, corresponding to the TPC inner containment vessel (interval 8) and the silicon pixels plus thermal shield (interval 2), respectively. The two sets of calibration weights, ${\Omega}_i$  and $\omega_i$, are very similar to each other, differing by only about ~2.5\%. On the other hand, one observes that for $R<$ 55 cm  the  uncertainties of ${\Omega}_i$ are smaller than for $\omega_i$, while they are larger for $R>$ 55 cm. The reason is that in the case of $\omega_i$ the uncertainties in the gas add to the corresponding ones in the given interval for $R<$ 55 cm because the mean momentum of the reconstructed photons in radial intervals "gas" and "i" are different, while, as the radial 
interval comes closer to the one in the gas ($R>$ 55 cm), part of the uncertainties cancels out. 
According to 
\hyperref[eq:OmegaToomegaratio]{Eq.~(\ref*{eq:OmegaToomegaratio})}, a difference in the ratio can be attributed to small differences in the reconstruction efficiency ratio ($\epsilon_{\gamma,\mathrm{gas}}/\epsilon_{\mathrm{track}}$) in RD with respect to the one obtained in the MC simulations, or even differences in the reference calibration material ($P_{\mathrm{gas}}$, see \hyperref[eq:OmegaToomegaratioWithP]{Eq.~(\ref*{eq:OmegaToomegaratioWithP})}). As a small difference of 2.5\% is observed,
the $\omega_i$ correction factors cannot be taken directly as material budget correction 
(see \hyperref[sec:Omega2omega]{Sec.~\ref*{sec:Omega2omega}} and \hyperref[eq:OmegaToomegaratio]{Eq.~(\ref*{eq:OmegaToomegaratio})}). Consequently, the values of ${\Omega}_i$ as shown in 
\hyperref[fig:TildeOmega]{ Fig.~\ref*{fig:TildeOmega}} and given in \hyperref[Tab:TildeOmega]{Table~\ref*{Tab:TildeOmega}} are taken as the best correction factors.

\begin{table}[hbt]
\centering 
\caption{Correction factors  ${\Omega}_i$
 for each radial interval used in this analysis, as well as the average value. Statistical, systematic, and total uncertainties are also given, except for the first two intervals where only statistical uncertainties are quoted. }
\begin{tabular}{|r|c| l | l| l | l|}
 \hline
$R$ interval & $R$ range~(cm)&  ${\Omega}_i$  & $\sigma_{\rm stat}$ \%& $\sigma_{\rm sys}$ \%&    $\sigma_{\rm total}$ \%    \\\hline\hline
0 &0--1.5 &   0.9859 & 1.2 & - &   - \\  
1 & 1.5--5 &   1.177 & 0.42 & -  &  -  \\ \hline
2 & 5--8.5 &   1.240 & 0.36 & 2.7  &    2.8  \\
3 & 8.5--13 &  1.238 & 0.42 & 0.77  &    0.9 \\
4 & 13--21 &   1.067 & 0.34 & 2.0  &    2.1 \\
5 & 21--33.5 & 1.081 & 0.25 & 1.7  &    1.7   \\
6 & 33.5--41 & 1.039 & 0.35 & 3.1  &   3.1 \\
7 & 41--55 &   1.001 & 0.30 & 1.5  &    1.5 \\
8 & 55--72 &  0.926 & 0.35 &  3.7  &   3.7   \\
9 & 72--95 &  0.943 & 0.19 & 3.7  &   3.7   \\
10 & 95--145 & 0.975 & 0.62 & 4.1  &   4.1  \\
11 &145--180 &0.932 & 0.89 & 1.4  &    1.6   \\ \hline\hline
 average  &  5--180 &1.04  & 0.312\%  & 2.5\%   &  2.5\% \\ \hline
 \end{tabular} 
 
\label{Tab:TildeOmega}
 \end{table}

\begin{figure}[htb]
    \centering
    \includegraphics[width=0.75\textwidth]{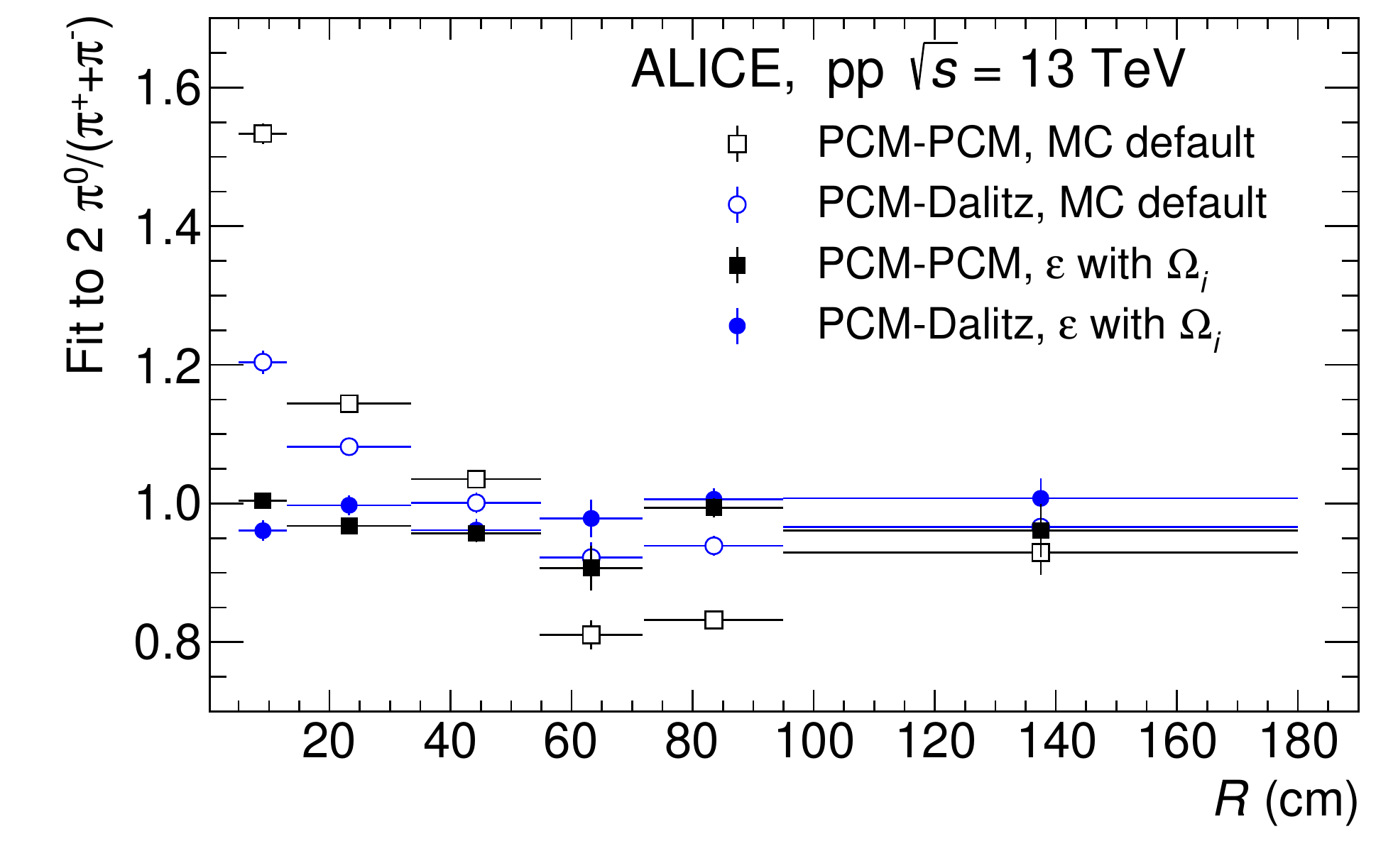}

    \caption{Zero-order polynomial fits to the ratio of neutral to charged pion transverse momentum distributions above 1 \GeVc when the two photons (PCM-PCM, $\pi^0\rightarrow \gamma \gamma$) or one photon (PCM-Dalitz, $\pi^0\rightarrow e^+e^-\gamma $) are selected within the given radial interval. The open symbols are obtained with efficiency corrections using the default MC , while the full symbols are obtained when the $\Omega_i$ calibration factors are used to weight the efficiency 
    (see \hyperref[eq:efi]{Eq.~\ref*{eq:efi}}).}
    \label{fig:pi0TopiChRBins}
\end{figure}

\begin{figure}[hbt]
\centering
\includegraphics[width=0.75\textwidth]{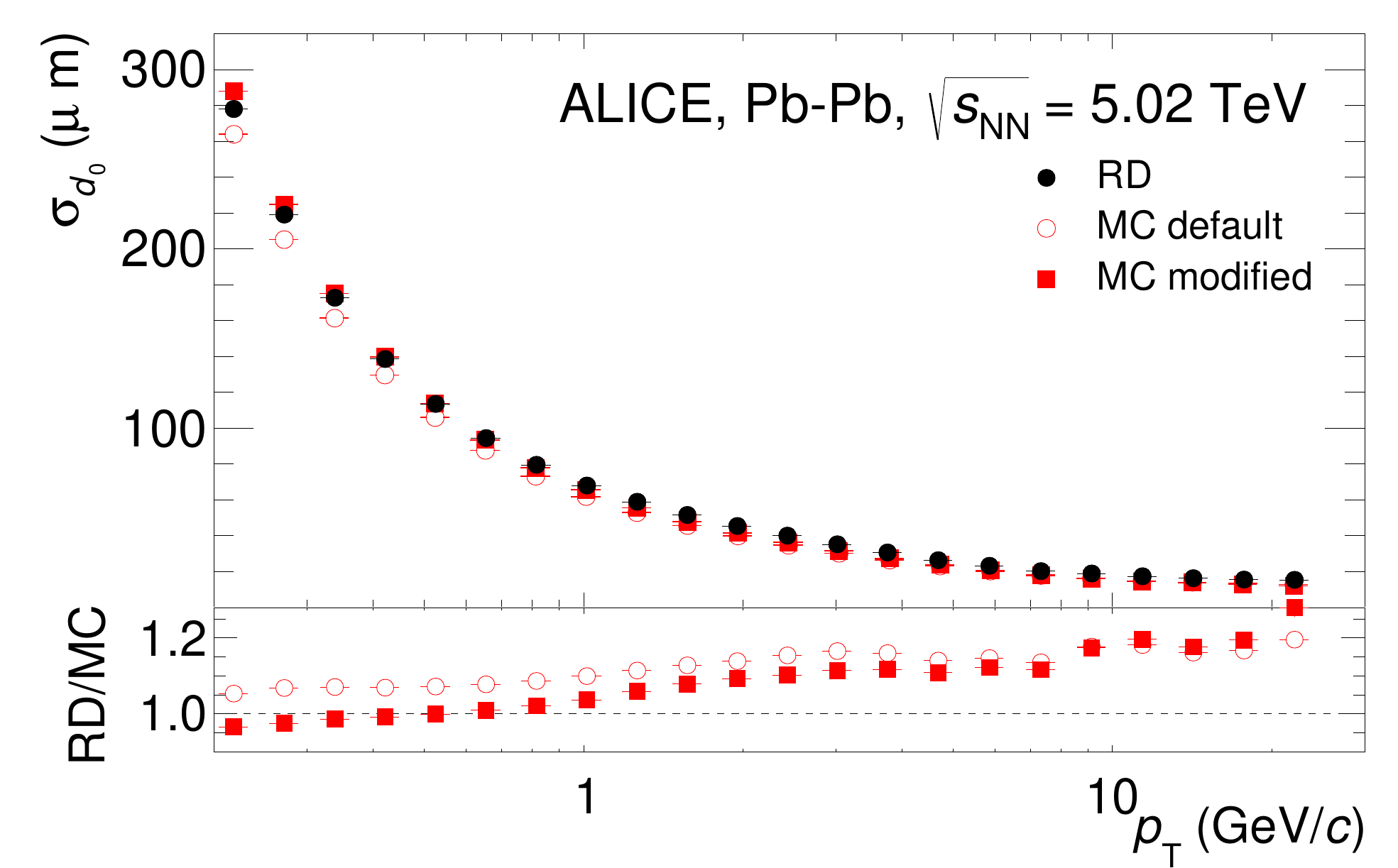}
\caption{Top: Impact-parameter resolution of reconstructed charged particles requiring a hit in the first ITS pixel layer as a function of \pT in RD, default MC, and modified MC with the correction factors as given in \hyperref[Tab:TildeOmega]{Table~\ref*{Tab:TildeOmega}}. Bottom: Ratio of 
impact-parameter resolution to the one in the default MC (open circles) and modified MC (squares).}
\label{fig:resD0}
\end{figure}

In order to further verify the correctness of the ${\Omega}_i$ values, \piz measurements~\cite{ALICE:2012wos,ALICE:2017ryd,ALICE:2018vhm,ALICE:2018mdl} are carried out selecting photons in a given radial interval at a time, before and after applying the correction factors. The transverse momentum spectra of \piz should not depend on the radial interval where the photons are reconstructed. \piz transverse momentum spectra measured with two (or one) decay photons within one radial interval\footnote{For the Dalitz or the hybrid (PCM-EMC or PCM-PHOS) reconstruction methods, the selection of the radial interval only applies to the PCM photon.} were analysed and compared to the spectra of charged pions~\cite{ALICE:2020jsh}, by fitting their ratios with a constant (see \hyperref[fig:pi0TopiChRBins]{ Fig.~\ref*{fig:pi0TopiChRBins}}). The dispersion (rms) of the fit results is large when using the default MC. 
The fit results show clearly the  same pattern as the calibration weights for the  PCM-Dalitz analysis  while for the PCM-PCM analysis the expected quadratic effect is observed.
When using the ${\Omega}_i$, the rms reduces by almost a factor 10 when both photons are reconstructed with the PCM method, or by a factor $\sim$ 4 when only one PCM photon is used in the reconstruction, i.e., reconstructing either the Dalitz decay or reconstructing the second photon with a calorimeter. 
Furthermore, the ratio of the \piz measurement in the complete radial range to the charged-pion measurement 
is in good agreement with the PYTHIA 8 expectations within one standard deviation.

Another observation corroborating the need for material budget 
correction factors is that the impact-parameter resolution of charged particles, i.e., the resolution of the reconstructed distance of closest approach of a track to the primary vertex, in RD is underestimated by the default MC. The small difference between data and MC of the particle composition plays a negligible effect, as most of the charged particles 
are charged pions.
\hyperref[fig:resD0]{ Fig.~\ref*{fig:resD0}}
shows the impact-parameter resolution of reconstructed charged particles with a hit in the first ITS pixel layer as a function of \pT for RD and for the default MC. The difference between RD and MC is usually corrected with an ad-hoc smearing of the track parameters in the MC.
On the other hand, a modified MC simulation where the correction factors as given in \hyperref[Tab:TildeOmega]{Table~\ref*{Tab:TildeOmega}} are used to scale the density of the detector materials reproduces the measured resolution for \pT $<$ 1 \GeVc, where the multiple scattering contribution is largest. This result confirms also that the assumption of attributing the correction factors to the material budget and not to the efficiency is correct. This demonstrates the importance of the material budget correction well beyond the reconstruction of photons with the conversion method.

\section{Conclusions}
\label{sec:con}
Two data-driven calibration methods of the detector material description in ALICE, one based on the precise knowledge of the ALICE TPC gas and
 the other based on the approximate pion isospin symmetry and named $\omega_i$ and $\Omega_i$, were developed. 
A reduction of the systematic uncertainty of almost a factor of two is achieved in the material budget up to a radius $R=$~180~cm corresponding approximately to the radial center of 
the TPC. Moreover, the differences between the description of the material distribution used for MC and the reality in the individual $R$ intervals are mitigated. This addresses the largest source of systematic uncertainty in analyses using photon conversions. It also reduces an important, and sometimes dominant, source of systematic uncertainty in analyses based on charged-particle tracking, in particular when secondary vertices are used. 
The upgraded ALICE experiment for Run 3~\cite{ALICE:2023udb} will continue to use these two calibration methods. Moreover, to assist the $\omega_i$ method, two calibrated tungsten wires were inserted in the inner and the outer barrels of ITS2~\cite{ALICE:2013nwm}.  
These two methods are general in nature and could be applied to any experiment in a similar fashion.


\newenvironment{acknowledgement}{\relax}{\relax}
\begin{acknowledgement}
\section*{Acknowledgements}

The ALICE Collaboration would like to thank all its engineers and technicians for their invaluable contributions to the construction of the experiment and the CERN accelerator teams for the outstanding performance of the LHC complex.
The ALICE Collaboration gratefully acknowledges the resources and support provided by all Grid centres and the Worldwide LHC Computing Grid (WLCG) collaboration.
The ALICE Collaboration acknowledges the following funding agencies for their support in building and running the ALICE detector:
A. I. Alikhanyan National Science Laboratory (Yerevan Physics Institute) Foundation (ANSL), State Committee of Science and World Federation of Scientists (WFS), Armenia;
Austrian Academy of Sciences, Austrian Science Fund (FWF): [M 2467-N36] and Nationalstiftung f\"{u}r Forschung, Technologie und Entwicklung, Austria;
Ministry of Communications and High Technologies, National Nuclear Research Center, Azerbaijan;
Conselho Nacional de Desenvolvimento Cient\'{\i}fico e Tecnol\'{o}gico (CNPq), Financiadora de Estudos e Projetos (Finep), Funda\c{c}\~{a}o de Amparo \`{a} Pesquisa do Estado de S\~{a}o Paulo (FAPESP) and Universidade Federal do Rio Grande do Sul (UFRGS), Brazil;
Bulgarian Ministry of Education and Science, within the National Roadmap for Research Infrastructures 2020-2027 (object CERN), Bulgaria;
Ministry of Education of China (MOEC) , Ministry of Science \& Technology of China (MSTC) and National Natural Science Foundation of China (NSFC), China;
Ministry of Science and Education and Croatian Science Foundation, Croatia;
Centro de Aplicaciones Tecnol\'{o}gicas y Desarrollo Nuclear (CEADEN), Cubaenerg\'{\i}a, Cuba;
Ministry of Education, Youth and Sports of the Czech Republic, Czech Republic;
The Danish Council for Independent Research | Natural Sciences, the VILLUM FONDEN and Danish National Research Foundation (DNRF), Denmark;
Helsinki Institute of Physics (HIP), Finland;
Commissariat \`{a} l'Energie Atomique (CEA) and Institut National de Physique Nucl\'{e}aire et de Physique des Particules (IN2P3) and Centre National de la Recherche Scientifique (CNRS), France;
Bundesministerium f\"{u}r Bildung und Forschung (BMBF) and GSI Helmholtzzentrum f\"{u}r Schwerionenforschung GmbH, Germany;
General Secretariat for Research and Technology, Ministry of Education, Research and Religions, Greece;
National Research, Development and Innovation Office, Hungary;
Department of Atomic Energy Government of India (DAE), Department of Science and Technology, Government of India (DST), University Grants Commission, Government of India (UGC) and Council of Scientific and Industrial Research (CSIR), India;
National Research and Innovation Agency - BRIN, Indonesia;
Istituto Nazionale di Fisica Nucleare (INFN), Italy;
Japanese Ministry of Education, Culture, Sports, Science and Technology (MEXT) and Japan Society for the Promotion of Science (JSPS) KAKENHI, Japan;
Consejo Nacional de Ciencia (CONACYT) y Tecnolog\'{i}a, through Fondo de Cooperaci\'{o}n Internacional en Ciencia y Tecnolog\'{i}a (FONCICYT) and Direcci\'{o}n General de Asuntos del Personal Academico (DGAPA), Mexico;
Nederlandse Organisatie voor Wetenschappelijk Onderzoek (NWO), Netherlands;
The Research Council of Norway, Norway;
Commission on Science and Technology for Sustainable Development in the South (COMSATS), Pakistan;
Pontificia Universidad Cat\'{o}lica del Per\'{u}, Peru;
Ministry of Education and Science, National Science Centre and WUT ID-UB, Poland;
Korea Institute of Science and Technology Information and National Research Foundation of Korea (NRF), Republic of Korea;
Ministry of Education and Scientific Research, Institute of Atomic Physics, Ministry of Research and Innovation and Institute of Atomic Physics and University Politehnica of Bucharest, Romania;
Ministry of Education, Science, Research and Sport of the Slovak Republic, Slovakia;
National Research Foundation of South Africa, South Africa;
Swedish Research Council (VR) and Knut \& Alice Wallenberg Foundation (KAW), Sweden;
European Organization for Nuclear Research, Switzerland;
Suranaree University of Technology (SUT), National Science and Technology Development Agency (NSTDA), Thailand Science Research and Innovation (TSRI) and National Science, Research and Innovation Fund (NSRF), Thailand;
Turkish Energy, Nuclear and Mineral Research Agency (TENMAK), Turkey;
National Academy of  Sciences of Ukraine, Ukraine;
Science and Technology Facilities Council (STFC), United Kingdom;
National Science Foundation of the United States of America (NSF) and United States Department of Energy, Office of Nuclear Physics (DOE NP), United States of America.
In addition, individual groups or members have received support from:
European Research Council, Strong 2020 - Horizon 2020 (grant nos. 950692, 824093), European Union;
Academy of Finland (Center of Excellence in Quark Matter) (grant nos. 346327, 346328), Finland;
Programa de Apoyos para la Superaci\'{o}n del Personal Acad\'{e}mico, UNAM, Mexico.

\end{acknowledgement}

\bibliographystyle{utphys}   
\bibliography{bibliography}

\newpage
\appendix

%
%

\section{The ALICE Collaboration}
\label{app:collab}
\begin{flushleft} 
\small

S.~Acharya\,\orcidlink{0000-0002-9213-5329}\,$^{\rm 125}$, 
D.~Adamov\'{a}\,\orcidlink{0000-0002-0504-7428}\,$^{\rm 86}$, 
A.~Adler$^{\rm 69}$, 
G.~Aglieri Rinella\,\orcidlink{0000-0002-9611-3696}\,$^{\rm 32}$, 
M.~Agnello\,\orcidlink{0000-0002-0760-5075}\,$^{\rm 29}$, 
N.~Agrawal\,\orcidlink{0000-0003-0348-9836}\,$^{\rm 50}$, 
Z.~Ahammed\,\orcidlink{0000-0001-5241-7412}\,$^{\rm 132}$, 
S.~Ahmad\,\orcidlink{0000-0003-0497-5705}\,$^{\rm 15}$, 
S.U.~Ahn\,\orcidlink{0000-0001-8847-489X}\,$^{\rm 70}$, 
I.~Ahuja\,\orcidlink{0000-0002-4417-1392}\,$^{\rm 37}$
% ...

\section*{Affiliation Notes}

$^{\rm I}$ Deceased\\
$^{\rm II}$ Also at: Italian National Agency for New Technologies, Energy and Sustainable Economic Development (ENEA), Bologna, Italy\\
$^{\rm III}$ Also at: Department of Applied Physics, Aligarh Muslim University, Aligarh, India\\
$^{\rm IV}$ Also at: Institute of Theoretical Physics, University of Wroclaw, Poland\\
$^{\rm V}$ Also at: An institution covered by a cooperation agreement with CERN\\

\section*{Collaboration Institutes}

$^{1}$ A.I. Alikhanyan National Science Laboratory (Yerevan Physics Institute) Foundation, Yerevan, Armenia\\
$^{2}$ AGH University of Science and Technology, Cracow, Poland\\
$^{3}$ Bogolyubov Institute for Theoretical Physics, National Academy of Sciences of Ukraine, Kiev, Ukraine\\
$^{4}$ Bose Institute, Department of Physics  and Centre for Astroparticle Physics and Space Science (CAPSS), Kolkata, India\\
$^{5}$ California Polytechnic State University, San Luis Obispo, California, United States\\
$^{6}$ Central China Normal University, Wuhan, China\\
$^{7}$ Centro de Aplicaciones Tecnol\'{o}gicas y Desarrollo Nuclear (CEADEN), Havana, Cuba\\
$^{8}$ Centro de Investigaci\'{o}n y de Estudios Avanzados (CINVESTAV), Mexico City and M\'{e}rida, Mexico\\
$^{9}$ Chicago State University, Chicago, Illinois, United States\\
$^{10}$ China Institute of Atomic Energy, Beijing, China\\
$^{11}$ Chungbuk National University, Cheongju, Republic of Korea\\
$^{12}$ Comenius University Bratislava, Faculty of Mathematics, Physics and Informatics, Bratislava, Slovak Republic\\
$^{13}$ COMSATS University Islamabad, Islamabad, Pakistan\\
$^{14}$ Creighton University, Omaha, Nebraska, United States\\
$^{15}$ Department of Physics, Aligarh Muslim University, Aligarh, India\\
$^{16}$ Department of Physics, Pusan National University, Pusan, Republic of Korea\\
$^{17}$ Department of Physics, Sejong University, Seoul, Republic of Korea\\
$^{18}$ Department of Physics, University of California, Berkeley, California, United States\\
$^{19}$ Department of Physics, University of Oslo, Oslo, Norway\\
$^{20}$ Department of Physics and Technology, University of Bergen, Bergen, Norway\\
$^{21}$ Dipartimento di Fisica, Universit\`{a} di Pavia, Pavia, Italy\\
$^{22}$ Dipartimento di Fisica dell'Universit\`{a} and Sezione INFN, Cagliari, Italy\\
$^{23}$ Dipartimento di Fisica dell'Universit\`{a} and Sezione INFN, Trieste, Italy\\
$^{24}$ Dipartimento di Fisica dell'Universit\`{a} and Sezione INFN, Turin, Italy\\
$^{25}$ Dipartimento di Fisica e Astronomia dell'Universit\`{a} and Sezione INFN, Bologna, Italy\\
$^{26}$ Dipartimento di Fisica e Astronomia dell'Universit\`{a} and Sezione INFN, Catania, Italy\\
$^{27}$ Dipartimento di Fisica e Astronomia dell'Universit\`{a} and Sezione INFN, Padova, Italy\\
$^{28}$ Dipartimento di Fisica `E.R.~Caianiello' dell'Universit\`{a} and Gruppo Collegato INFN, Salerno, Italy\\
$^{29}$ Dipartimento DISAT del Politecnico and Sezione INFN, Turin, Italy\\
$^{30}$ Dipartimento di Scienze MIFT, Universit\`{a} di Messina, Messina, Italy\\
$^{31}$ Dipartimento Interateneo di Fisica `M.~Merlin' and Sezione INFN, Bari, Italy\\
$^{32}$ European Organization for Nuclear Research (CERN), Geneva, Switzerland\\
$^{33}$ Faculty of Electrical Engineering, Mechanical Engineering and Naval Architecture, University of Split, Split, Croatia\\
$^{34}$ Faculty of Engineering and Science, Western Norway University of Applied Sciences, Bergen, Norway\\
$^{35}$ Faculty of Nuclear Sciences and Physical Engineering, Czech Technical University in Prague, Prague, Czech Republic\\
$^{36}$ Faculty of Physics, Sofia University, Sofia, Bulgaria\\
$^{37}$ Faculty of Science, P.J.~\v{S}af\'{a}rik University, Ko\v{s}ice, Slovak Republic\\
$^{38}$ Frankfurt Institute for Advanced Studies, Johann Wolfgang Goethe-Universit\"{a}t Frankfurt, Frankfurt, Germany\\
$^{39}$ Fudan University, Shanghai, China\\
$^{40}$ Gangneung-Wonju National University, Gangneung, Republic of Korea\\
$^{41}$ Gauhati University, Department of Physics, Guwahati, India\\
$^{42}$ Helmholtz-Institut f\"{u}r Strahlen- und Kernphysik, Rheinische Friedrich-Wilhelms-Universit\"{a}t Bonn, Bonn, Germany\\
$^{43}$ Helsinki Institute of Physics (HIP), Helsinki, Finland\\
$^{44}$ High Energy Physics Group,  Universidad Aut\'{o}noma de Puebla, Puebla, Mexico\\
$^{45}$ Horia Hulubei National Institute of Physics and Nuclear Engineering, Bucharest, Romania\\
$^{46}$ Indian Institute of Technology Bombay (IIT), Mumbai, India\\
$^{47}$ Indian Institute of Technology Indore, Indore, India\\
$^{48}$ INFN, Laboratori Nazionali di Frascati, Frascati, Italy\\
$^{49}$ INFN, Sezione di Bari, Bari, Italy\\
$^{50}$ INFN, Sezione di Bologna, Bologna, Italy\\
$^{51}$ INFN, Sezione di Cagliari, Cagliari, Italy\\
$^{52}$ INFN, Sezione di Catania, Catania, Italy\\
$^{53}$ INFN, Sezione di Padova, Padova, Italy\\
$^{54}$ INFN, Sezione di Pavia, Pavia, Italy\\
$^{55}$ INFN, Sezione di Torino, Turin, Italy\\
$^{56}$ INFN, Sezione di Trieste, Trieste, Italy\\
$^{57}$ Inha University, Incheon, Republic of Korea\\
$^{58}$ Institute for Gravitational and Subatomic Physics (GRASP), Utrecht University/Nikhef, Utrecht, Netherlands\\
$^{59}$ Institute of Experimental Physics, Slovak Academy of Sciences, Ko\v{s}ice, Slovak Republic\\
$^{60}$ Institute of Physics, Homi Bhabha National Institute, Bhubaneswar, India\\
$^{61}$ Institute of Physics of the Czech Academy of Sciences, Prague, Czech Republic\\
$^{62}$ Institute of Space Science (ISS), Bucharest, Romania\\
$^{63}$ Institut f\"{u}r Kernphysik, Johann Wolfgang Goethe-Universit\"{a}t Frankfurt, Frankfurt, Germany\\
$^{64}$ Instituto de Ciencias Nucleares, Universidad Nacional Aut\'{o}noma de M\'{e}xico, Mexico City, Mexico\\
$^{65}$ Instituto de F\'{i}sica, Universidade Federal do Rio Grande do Sul (UFRGS), Porto Alegre, Brazil\\
$^{66}$ Instituto de F\'{\i}sica, Universidad Nacional Aut\'{o}noma de M\'{e}xico, Mexico City, Mexico\\
$^{67}$ iThemba LABS, National Research Foundation, Somerset West, South Africa\\
$^{68}$ Jeonbuk National University, Jeonju, Republic of Korea\\
$^{69}$ Johann-Wolfgang-Goethe Universit\"{a}t Frankfurt Institut f\"{u}r Informatik, Fachbereich Informatik und Mathematik, Frankfurt, Germany\\
$^{70}$ Korea Institute of Science and Technology Information, Daejeon, Republic of Korea\\
$^{71}$ KTO Karatay University, Konya, Turkey\\
$^{72}$ Laboratoire de Physique des 2 Infinis, Ir\`{e}ne Joliot-Curie, Orsay, France\\
$^{73}$ Laboratoire de Physique Subatomique et de Cosmologie, Universit\'{e} Grenoble-Alpes, CNRS-IN2P3, Grenoble, France\\
$^{74}$ Lawrence Berkeley National Laboratory, Berkeley, California, United States\\
$^{75}$ Lund University Department of Physics, Division of Particle Physics, Lund, Sweden\\
$^{76}$ Nagasaki Institute of Applied Science, Nagasaki, Japan\\
$^{77}$ Nara Women{'}s University (NWU), Nara, Japan\\
$^{78}$ National and Kapodistrian University of Athens, School of Science, Department of Physics , Athens, Greece\\
$^{79}$ National Centre for Nuclear Research, Warsaw, Poland\\
$^{80}$ National Institute of Science Education and Research, Homi Bhabha National Institute, Jatni, India\\
$^{81}$ National Nuclear Research Center, Baku, Azerbaijan\\
$^{82}$ National Research and Innovation Agency - BRIN, Jakarta, Indonesia\\
$^{83}$ Niels Bohr Institute, University of Copenhagen, Copenhagen, Denmark\\
$^{84}$ Nikhef, National institute for subatomic physics, Amsterdam, Netherlands\\
$^{85}$ Nuclear Physics Group, STFC Daresbury Laboratory, Daresbury, United Kingdom\\
$^{86}$ Nuclear Physics Institute of the Czech Academy of Sciences, Husinec-\v{R}e\v{z}, Czech Republic\\
$^{87}$ Oak Ridge National Laboratory, Oak Ridge, Tennessee, United States\\
$^{88}$ Ohio State University, Columbus, Ohio, United States\\
$^{89}$ Physics department, Faculty of science, University of Zagreb, Zagreb, Croatia\\
$^{90}$ Physics Department, Panjab University, Chandigarh, India\\
$^{91}$ Physics Department, University of Jammu, Jammu, India\\
$^{92}$ Physics Program and International Institute for Sustainability with Knotted Chiral Meta Matter (SKCM2), Hiroshima University, Hiroshima, Japan\\
$^{93}$ Physikalisches Institut, Eberhard-Karls-Universit\"{a}t T\"{u}bingen, T\"{u}bingen, Germany\\
$^{94}$ Physikalisches Institut, Ruprecht-Karls-Universit\"{a}t Heidelberg, Heidelberg, Germany\\
$^{95}$ Physik Department, Technische Universit\"{a}t M\"{u}nchen, Munich, Germany\\
$^{96}$ Politecnico di Bari and Sezione INFN, Bari, Italy\\
$^{97}$ Research Division and ExtreMe Matter Institute EMMI, GSI Helmholtzzentrum f\"ur Schwerionenforschung GmbH, Darmstadt, Germany\\
$^{98}$ Saga University, Saga, Japan\\
$^{99}$ Saha Institute of Nuclear Physics, Homi Bhabha National Institute, Kolkata, India\\
$^{100}$ School of Physics and Astronomy, University of Birmingham, Birmingham, United Kingdom\\
$^{101}$ Secci\'{o}n F\'{\i}sica, Departamento de Ciencias, Pontificia Universidad Cat\'{o}lica del Per\'{u}, Lima, Peru\\
$^{102}$ Stefan Meyer Institut f\"{u}r Subatomare Physik (SMI), Vienna, Austria\\
$^{103}$ SUBATECH, IMT Atlantique, Nantes Universit\'{e}, CNRS-IN2P3, Nantes, France\\
$^{104}$ Sungkyunkwan University, Suwon City, Republic of Korea\\
$^{105}$ Suranaree University of Technology, Nakhon Ratchasima, Thailand\\
$^{106}$ Technical University of Ko\v{s}ice, Ko\v{s}ice, Slovak Republic\\
$^{107}$ The Henryk Niewodniczanski Institute of Nuclear Physics, Polish Academy of Sciences, Cracow, Poland\\
$^{108}$ The University of Texas at Austin, Austin, Texas, United States\\
$^{109}$ Universidad Aut\'{o}noma de Sinaloa, Culiac\'{a}n, Mexico\\
$^{110}$ Universidade de S\~{a}o Paulo (USP), S\~{a}o Paulo, Brazil\\
$^{111}$ Universidade Estadual de Campinas (UNICAMP), Campinas, Brazil\\
$^{112}$ Universidade Federal do ABC, Santo Andre, Brazil\\
$^{113}$ University of Cape Town, Cape Town, South Africa\\
$^{114}$ University of Houston, Houston, Texas, United States\\
$^{115}$ University of Jyv\"{a}skyl\"{a}, Jyv\"{a}skyl\"{a}, Finland\\
$^{116}$ University of Kansas, Lawrence, Kansas, United States\\
$^{117}$ University of Liverpool, Liverpool, United Kingdom\\
$^{118}$ University of Science and Technology of China, Hefei, China\\
$^{119}$ University of South-Eastern Norway, Kongsberg, Norway\\
$^{120}$ University of Tennessee, Knoxville, Tennessee, United States\\
$^{121}$ University of the Witwatersrand, Johannesburg, South Africa\\
$^{122}$ University of Tokyo, Tokyo, Japan\\
$^{123}$ University of Tsukuba, Tsukuba, Japan\\
$^{124}$ University Politehnica of Bucharest, Bucharest, Romania\\
$^{125}$ Universit\'{e} Clermont Auvergne, CNRS/IN2P3, LPC, Clermont-Ferrand, France\\
$^{126}$ Universit\'{e} de Lyon, CNRS/IN2P3, Institut de Physique des 2 Infinis de Lyon, Lyon, France\\
$^{127}$ Universit\'{e} de Strasbourg, CNRS, IPHC UMR 7178, F-67000 Strasbourg, France, Strasbourg, France\\
$^{128}$ Universit\'{e} Paris-Saclay Centre d'Etudes de Saclay (CEA), IRFU, D\'{e}partment de Physique Nucl\'{e}aire (DPhN), Saclay, France\\
$^{129}$ Universit\`{a} degli Studi di Foggia, Foggia, Italy\\
$^{130}$ Universit\`{a} del Piemonte Orientale, Vercelli, Italy\\
$^{131}$ Universit\`{a} di Brescia, Brescia, Italy\\
$^{132}$ Variable Energy Cyclotron Centre, Homi Bhabha National Institute, Kolkata, India\\
$^{133}$ Warsaw University of Technology, Warsaw, Poland\\
$^{134}$ Wayne State University, Detroit, Michigan, United States\\
$^{135}$ Westf\"{a}lische Wilhelms-Universit\"{a}t M\"{u}nster, Institut f\"{u}r Kernphysik, M\"{u}nster, Germany\\
$^{136}$ Wigner Research Centre for Physics, Budapest, Hungary\\
$^{137}$ Yale University, New Haven, Connecticut, United States\\
$^{138}$ Yonsei University, Seoul, Republic of Korea\\
$^{139}$  Zentrum  f\"{u}r Technologie und Transfer (ZTT), Worms, Germany\\
$^{140}$ Affiliated with an institute covered by a cooperation agreement with CERN\\
$^{141}$ Affiliated with an international laboratory covered by a cooperation agreement with CERN.\\

\end{flushleft}

\end{document}